\documentclass[acmsmall,nonacm]{acmart}
\usepackage{graphicx}
\usepackage{amsmath}
\usepackage{appendix}
\usepackage{subcaption}
\usepackage{caption}
\usepackage{tabularx}
\usepackage{braket}
\usepackage{pifont}
\usepackage{siunitx}
\usepackage{xcolor}
\usepackage{comment} 
 
\newcommand{\qcomm}{{\textsf{qcomm}}}  % Sans-serif
\newcommand{\cone}{\ding{182}}
\newcommand{\ctwo}{\ding{183}}
\newcommand{\cthree}{\ding{184}}
\newcommand{\cfour}{\ding{185}}
\newcommand{\cfive}{\ding{186}}

\AtBeginDocument{%
  }

\begin{document}

\title{Assessing the Role of Communication in Modular Multi-Core Quantum Systems}
\thanks{This work is a significantly extended version of the paper entitled 
“Assessing the Role of Communication in Scalable Multi-Core Quantum Architectures,” 
published in the Proceedings of the IEEE 17th International Symposium on Embedded Multicore/Many-core Systems-on-Chip (MCSoC), 2024.}

\author{Maurizio Palesi}
\email{maurizio.palesi@unict.it}
\author{Enrico Russo}
\email{enrico.russo@phd.unict.it}
\author{Giuseppe Ascia}
\email{giuseppe.ascia@unict.it}
\author{Hamaad Rafique}
\email{hamaad.rafique@phd.unict.it}
\author{Davide Patti}
\email{davide.patti@unict.it}
\author{Vincenzo Catania}
\email{vincenzo.catania@unict.it}
\email{first.last@unict.it}
\affiliation{%
  \institution{Università degli Studi di Catania}
  \city{Catania}
  \country{Italy}
}

\author{Sergi Abadal}
\email{sergi.abadal@upc.edu}
\author{Abhijit Das}
\email{abhijit.das@upc.edu}
\author{Pau Escofet}
\email{pau.escofet@upc.edu}
\author{Eduard Alarcon}
\email{eduard.alarcon@upc.edu}
\affiliation{%
  \institution{Universitat Politècnica de Catalunya}
  \city{Barcelona}
  \country{Spain}
}

\author{Carmen G. Almudéver}
\email{Carmen G. Almudéver}
\affiliation{%
  \institution{Universitat Politècnica de València}
  \city{Valencia}
  \country{Spain}
}

\renewcommand{\shortauthors}{Maurizio Palesi et al.}

\renewcommand{\shortauthors}{Palesi et al.}

\begin{abstract}
The scalability of quantum computing is constrained by the physical and architectural limitations of monolithic quantum processors. Modular multi-core quantum architectures, which interconnect multiple quantum cores (QCs) via classical and quantum-coherent links, offer a promising alternative to address these challenges. However, transitioning to a modular architecture introduces communication overhead, where classical communication plays a crucial role in executing quantum algorithms by transmitting measurement outcomes and synchronizing operations across QCs. Understanding the impact of classical communication on execution time is therefore essential for optimizing system performance.

In this work, we introduce \qcomm, an open-source simulator designed to evaluate the role of classical communication in modular quantum computing architectures. \qcomm{} provides a high-level execution and timing model that captures the interplay between quantum gate execution, entanglement distribution, teleportation protocols, and classical communication latency. We conduct an extensive experimental analysis to quantify the impact of classical communication bandwidth, interconnect types, and quantum circuit mapping strategies on overall execution time. Furthermore, we assess classical communication overhead when executing real quantum benchmarks mapped onto a cryogenically-controlled multi-core quantum system. Our results show that, while classical communication is generally not the dominant contributor to execution time, its impact becomes increasingly relevant in optimized scenarios---such as improved quantum technology, large-scale interconnects, or communication-aware circuit mappings. These findings provide useful insights for the design of scalable modular quantum architectures and highlight the importance of evaluating classical communication as a performance-limiting factor in future systems.
\end{abstract}

\begin{CCSXML}
<ccs2012>
   <concept>
       <concept_id>10010520.10010521.10010542.10010550</concept_id>
       <concept_desc>Computer systems organization~Quantum computing</concept_desc>
       <concept_significance>500</concept_significance>
       </concept>
   <concept>
       <concept_id>10003033.10003079.10003081</concept_id>
       <concept_desc>Networks~Network simulations</concept_desc>
       <concept_significance>500</concept_significance>
       </concept>
   <concept>
       <concept_id>10003752.10003753.10010622</concept_id>
       <concept_desc>Theory of computation~Abstract machines</concept_desc>
       <concept_significance>300</concept_significance>
       </concept>
 </ccs2012>
\end{CCSXML}

\ccsdesc[500]{Computer systems organization~Quantum computing}
\ccsdesc[500]{Networks~Network simulations}
\ccsdesc[300]{Theory of computation~Abstract machines}

\keywords{Quantum Computing, Modular Quantum Architectures, Quantum Communication,Classical Communication Overhead, Network-on-Chip (NoC), Wireless Network-on-Chip (WiNoC), Quantum Core Interconnection, Scalability in Quantum Systems, Quantum Simulation Tools, Multi-Core Quantum Processors}

\maketitle

\begin{acks}
Authors gratefully acknowledge funding from the European Commission through HORIZON-EIC-2022-PATHFINDEROPEN01-101099697 (QUADRATURE)
\end{acks}

\section{Introduction}

Quantum computing has emerged as a revolutionary paradigm, leveraging the principles of quantum mechanics to perform computations that are infeasible for classical systems~\cite{nielsen_chuang_2010,arute_nature19}.
However, many useful quantum algorithms---such as those for factoring, search, or quantum simulation---require a large number of qubits, far beyond what current devices can support. This demand is not only driven by algorithmic complexity but also by the need for quantum error correction (QEC), which introduces significant qubit overhead to protect fragile quantum information from decoherence and gate errors~\cite{fowler_phreview12}.

Current quantum hardware falls into the category of Noisy Intermediate-Scale Quantum (NISQ) systems, characterized by tens to hundreds of qubits, relatively high error rates, and the absence of full-scale QEC. While NISQ devices have enabled early experimental demonstrations, achieving fault-tolerant and scalable quantum computation will require architectures that can support thousands or even millions of physical qubits~\cite{preskill_quantum18}. As research progresses toward this goal, modular multi-core architectures have gained attention as a promising alternative to monolithic quantum processors~\cite{jnan_phr22,alarcorn_iscas23}. Unlike monolithic designs, which face significant scalability limitations due to control circuit integration, wiring complexity, and increased qubit interactions, modular architectures interconnect multiple quantum cores (QCs) via classical and quantum-coherent links~\cite{rached_arxiv25}. This approach enhances scalability while maintaining quantum coherence. However, transitioning from a monolithic to a multi-core quantum architecture introduces new challenges, particularly in communication overhead, both quantum and classical.

As quantum hardware advances, we can envision an architectural evolution from monolithic quantum processors controlled at room temperature, to room-temperature-controlled multi-core quantum systems, and eventually to fully cryogenically-controlled multi-core architectures~\cite{alarcorn_iscas23,almudever_date24}. While the first two stages are extensions of current experimental platforms, the final stage represents a significant shift in design: cryogenic control logic integrated closely with quantum cores to minimize latency and improve fidelity. This emerging class of architectures, which is not yet fully defined in the literature, introduces new challenges in system-level organization, control distribution, and communication infrastructure.

\begin{comment}
%%% LONG VERSION
    In this context, a key aspect of modular architectures is the ability to enable quantum communication between physically separated QCs. Multiple architectures and methodologies have been proposed to address this challenge, which can be broadly categorized into communication protocols and physical technologies. Protocols define the logical mechanism by which quantum information is exchanged across QCs. Common approaches include quantum teleportation~\cite{bose_prl03}---also referred to as \emph{teledata}---which transmits a quantum state via entanglement and classical communication~\cite{gottesman1999quantum}, direct quantum state transfer, and remote gate execution~\cite{vanmeterqic06}---known as \emph{telegate}---in which operations are applied to remote qubits without moving their quantum states. On the other hand, physical technologies implement these protocols through hardware mechanisms such as ion shuttling~\cite{hensinger_apl06}, cavity-mediated photonic links~\cite{ritter_nature12}, and superconducting microwave buses~\cite{sillanpaa_nature07}, each offering distinct trade-offs in fidelity, scalability, and connectivity.
\end{comment}

In this context, a key aspect of modular architectures is the ability to enable quantum communication between physically separated QCs. Multiple architectures and methodologies have been proposed to address this challenge, which can be broadly categorized into communication protocols and physical technologies. Protocols define the logical mechanism by which quantum information is exchanged across QCs. Common approaches include quantum teleportation~\cite{bose_prl03}, direct quantum state transfer, and remote gate execution~\cite{vanmeterqic06}. On the other hand, physical technologies implement these protocols through hardware mechanisms such as ion shuttling~\cite{hensinger_apl06}, cavity-mediated photonic links~\cite{ritter_nature12}, and superconducting microwave buses~\cite{sillanpaa_nature07}, each offering distinct trade-offs in fidelity, scalability, and connectivity.
    
\begin{comment}
    In this work, we focus on quantum teleportation as the underlying communication protocol, due to its compatibility with modular quantum architectures and its decoupling of entanglement distribution from data transmission. Since teleportation requires the exchange of classical information between communicating QCs, we model a Network-on-Chip (NoC)~\cite{benini_computer02} as the classical communication fabric. The NoC is used not only to support the teleportation protocol but also to handle the transmission of control, data, and synchronization signals throughout the system.

    Efficient communication between QCs is critical to the overall performance of a modular quantum system~\cite{das_jestcs24}, as quantum algorithms often require frequent inter-core qubit operations. While quantum communication enables entanglement-based state transfer via quantum teleportation, classical communication remains an essential component of the process, as it is required to transmit measurement outcomes and synchronize operations between QCs. Consequently, understanding the impact of classical communication on execution time is crucial for optimizing modular quantum architectures.
\end{comment}
In this work, we focus on quantum teleportation as the underlying communication protocol, due to its compatibility with modular quantum architectures and its decoupling of entanglement distribution from data transmission. Since teleportation requires the exchange of classical information between communicating QCs, we model a Network-on-Chip (NoC)~\cite{benini_computer02} as the classical communication fabric. The NoC is used not only to support the teleportation protocol but also to handle the transmission of control, data, and synchronization signals throughout the system.

Given the central role of classical communication in enabling teleportation and coordinating inter-core operations, analyzing its impact on execution time is essential~\cite{escofet_qce25}. In modular quantum systems, where quantum algorithms involve frequent inter-core interactions, the performance of the classical communication layer can significantly influence overall system efficiency. This motivates our exploration of NoC-based communication strategies and their effect on quantum workload execution.

To support the exploration of these next-generation systems, we introduce \qcomm{}\footnote{\url{https://github.com/mpalesi/qcomm}.}, an open-source simulator specifically designed for modular, cryogenically-controlled quantum architectures~\cite{palesi_mcsoc24}. \qcomm{} provides a high-level architectural abstraction that models the interaction between quantum gate execution, entanglement generation and distribution, teleportation protocols, and classical communication latency. It enables design space exploration by allowing researchers to evaluate how architectural and micro-architectural parameters---such as interconnect topology, bandwidth, and quantum core configurations---impact overall system performance, with a particular focus on communication-driven overheads.

Existing quantum computing simulators, such as Qiskit Aer~\cite{qiskit2024}, Azure Quantum~\cite{azure_quantum}, and the MATLAB Support Package for Quantum Computing~\cite{matlab_quantum}, primarily focus on simulating quantum circuit execution on a single monolithic processor. While these tools are highly effective for algorithm validation and small-scale quantum systems, they do not support the modeling of distributed architectures, where multiple quantum processing units communicate via quantum and classical channels. On the other end of the spectrum, quantum network simulators such as NetSquid~\cite{coopmans_cp21}, SeQUeNCe~\cite{wu_qst21, wu_tomacs24}, SimulaQron~\cite{dahlberg_qst18}, SQUANCH~\cite{squanch2018}, and QuNetSim~\cite{diadamo_tqe21} enable the study of quantum communication protocols, network behaviors, and entanglement distribution at the physical or protocol level. However, they are not designed to simulate quantum algorithm execution on modular multi-core processors or to evaluate architectural trade-offs at the system level. To bridge this gap, \qcomm{} offers a unique simulation environment that complements both circuit-level and network-level tools. By modeling high-level execution time and communication interactions in modular architectures, it provides a valuable platform for researchers exploring the architectural challenges of scalable, distributed quantum computing.

\begin{table}
    \centering
    \caption{Comparison of \qcomm{} with existing quantum simulators.}
    \label{tab:sim_comparison}
    \small
    \begin{tabularx}{\linewidth}{XXXXXXX} % {lcccccc}
    \toprule
    Simulator & Abstraction Level & Multi-core / \newline Modularity & 
    Classical \newline Communication Modeling & Teleportation \newline Protocols & 
    Cryogenic \newline Control \newline Support \\
    \midrule
    Qiskit Aer~\cite{qiskit2024} & Gate/state level & No & No & No & No \\
    \midrule
    NetSquid~\cite{coopmans_cp21} & Quantum \newline network \newline \scriptsize (protocol-level) & Limited  \newline \scriptsize (nodes) & Yes  \newline \scriptsize (network stack) & Yes & No \\
    \midrule
    SeQUeNCe~\cite{wu_qst21} & Quantum \newline network  \newline \scriptsize (discrete-event) & Yes  \newline \scriptsize (network nodes) & Yes & Yes \newline \scriptsize (entanglement distribution, routing) & No \\
    \midrule
    SimulaQron~\cite{dahlberg_qst18} & Virtual quantum internet & Yes  \newline \scriptsize(virtual nodes) & Yes & Limited \newline \scriptsize (ideal teleport) & No \\
    \midrule
    QuNetSim~\cite{diadamo_tqe21} & Quantum network protocol & Yes  \newline \scriptsize (apps, nodes) & Yes & Yes & No \\
    \midrule
    qcomm  \newline \scriptsize (this work) & Architectural \newline \scriptsize (execution + \newline timing model) & 
    Yes \newline \scriptsize (multi-core QCs) & Yes \newline \scriptsize (NoC, WiNoC) & 
    Yes \newline \scriptsize (TPS/TPD variants) & Yes \newline \scriptsize (cryogenic CU \newline integration) \\
    \bottomrule
    \end{tabularx}
\end{table}
To better contextualize \qcomm{}, Table~\ref{tab:sim_comparison} compares it with existing simulation tools. While frameworks such as Qiskit Aer focus on gate-level state evolution, and network simulators such as NetSquid, SeQUeNCe, or SimulaQron emphasize quantum network protocols, none of these capture the architectural interplay between classical communication, teleportation protocols, and cryogenic control in modular multi-core systems. \qcomm{} fills this gap by operating at the architectural level, modeling execution time as a function of both quantum gate delays and classical communication parameters.

This paper makes the following key contributions:
\begin{itemize}
    \item We develop \qcomm, a modular simulation framework for evaluating classical communication in multi-core quantum architectures.  

    \item We provide a comprehensive execution model that incorporates quantum gate execution, teleportation, and classical communication delays.

    \item We propose a high-level, component-aware timing model tailored for modular quantum architectures, which captures the cost of inter-core communication and teleportation protocols. Unlike gate-level simulators, our model allows for parametric evaluation of architectural decisions without requiring low-level quantum state simulation.
    
    \item We perform an extensive experimental analysis to quantify the impact of interconnect options (wired vs. wireless NoC), classical communication bandwidth, and circuit-to-core mapping strategies on execution time.  

    \item We investigate the role of classical communication in real quantum benchmarks.      
\end{itemize}

The remainder of this paper is organized as follows. Section~\ref{sec:background} presents background concepts, including qubits, quantum gates, entanglement, and the teleportation protocol. Section~\ref{sec:reference_arch} describes the reference modular quantum architecture while Section~\ref{sec:timing_model} introduces the timing model used in our analysis. Section~\ref{sec:qcomm} details the \qcomm{} simulator, and Section~\ref{sec:experiments} presents experimental results and performance evaluations. Finally, Section~\ref{sec:conclusion} concludes the paper and outlines future research directions.

\section{Background}
\label{sec:background}
This section provides a brief and non-exhaustive overview of key quantum computing concepts that are essential for understanding the rest of the paper. It is intended as a minimal background and not as a comprehensive introduction to quantum information science. Readers interested in a deeper and more rigorous treatment of the subject are encouraged to consult standard references, such as~\cite{nielsen_chuang_2010}.

Quantum computing is based on the principles of quantum mechanics, which enable fundamentally new ways of processing information. Unlike classical computing, which relies on bits that take values of either 0 or 1, quantum computing leverages \emph{qubits} (quantum bits), which can exist in multiple states simultaneously due to the phenomenon of \emph{quantum superposition}.  Mathematically, a qubit's state is represented as:  
\[ \ket{\psi} = \alpha \ket{0} + \beta \ket{1} \]
where $\alpha$ and $\beta$ are complex probability amplitudes that satisfy $|\alpha|^2 + |\beta|^2 = 1$. This superposition property enables quantum computers to perform parallel computations and process vast amounts of information more efficiently than classical systems.  

A fundamental aspect of quantum mechanics is the process of \emph{measurement}. When a qubit in a superposition state is measured in the computational basis, its state collapses probabilistically to either $\ket{0}$ or $\ket{1}$, with probabilities $|\alpha|^2$ and $|\beta|^2$, respectively. This irreversible collapse prevents access to the full quantum state through measurement alone. 

Quantum computation is performed using \emph{quantum gates}, which manipulate qubits through unitary operations. These gates are the quantum analogs of classical logic gates but operate on superpositions, enabling complex transformations of quantum states. Common quantum gates include the Hadamard ($H$) gate, the Pauli gates ($X$, $Y$, $Z$), and the CNOT (Controlled-NOT) gate. The Hadamard gate, creates superposition by transforming $\ket{0}$ into $(\ket{0} + \ket{1}) / \sqrt{2}$. The Pauli gates apply specific rotations to qubit states. The CNOT gate is a two-qubit quantum gate that flips the state of the target qubit if the control qubit is in the $\ket{1}$ state.

A \emph{quantum circuit} is a sequence of quantum gates applied to a set of qubits to implement a given algorithm. Each gate acts on one or more qubits, with multi-qubit gates (such as CNOT) requiring specific connectivity between the involved qubits. In physical quantum computers, however, qubits are typically arranged according to a hardware-specific \emph{connectivity map}, which defines which pairs of qubits can interact directly. For a multi-qubit gate to be executed, the target qubits must be adjacent in this map. When they are not, a routing procedure is required to bring them together, usually by applying a series of SWAP operations that move the qubit states along the connectivity graph until the gate can be applied. This process introduces additional gates, thereby increasing the circuit depth. Greater circuit depth is problematic in current quantum technologies due to the limited coherence time of physical qubits---the short period during which they can maintain quantum information reliably. As depth increases, so does the risk of decoherence and error accumulation, which degrade the fidelity of the computation. Different quantum hardware platforms adopt different connectivity constraints. For instance, superconducting qubit platforms (such as those used by IBM and Google) often use 2D grid topologies, while trapped-ion systems offer more flexible connectivity. Because of these architectural differences, the tasks of qubit assignment (also called qubit mapping) and routing are critical steps in the compilation process~\cite{siraichi_cgo18,li_asplos19}.

\emph{Entanglement} is a uniquely quantum phenomenon where two or more qubits become correlated in such a way that the state of one qubit instantaneously influences the state of the other, regardless of distance. This property is crucial for quantum communication and quantum computing, as it enables powerful operations such as \emph{quantum teleportation} and \emph{superdense coding}.  

Another fundamental principle is the \emph{no-cloning theorem}, which states that it is impossible to create an exact copy of an arbitrary unknown quantum state. This result, unique to quantum information, arises from the linearity of quantum operations. Unlike classical bits, qubits cannot be duplicated without altering the original state. The no-cloning theorem underpins the need for quantum teleportation: since a quantum state cannot be copied and sent, it must be transferred via entanglement and classical communication.

\begin{figure}
    \centering
    \includegraphics[width=0.9\columnwidth]{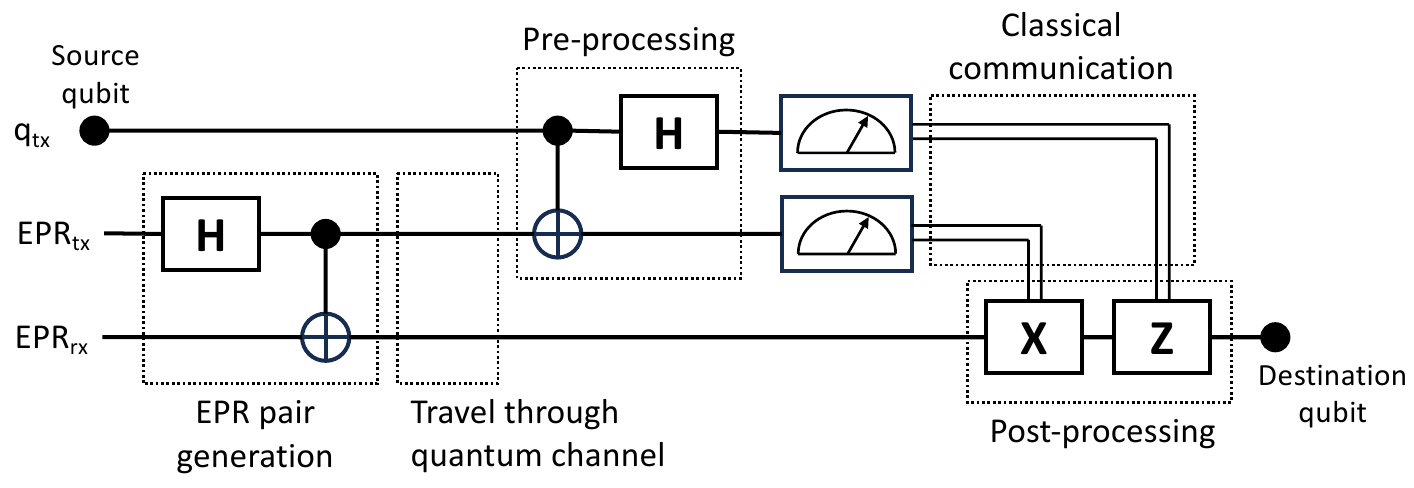}
    \caption{Steps involved in teleportation protocol.}
    \label{fig:teleportation}
\end{figure}
The \emph{teleportation protocol} is a method for transferring an unknown quantum state from one qubit to another without physically moving the qubit itself. It relies on entanglement and classical communication. The teleportation process consists of five sequential phases: EPR generation, EPR distribution, pre-processing, classical communication, and post-processing, as illustrated in Figure~\ref{fig:teleportation}. In the EPR generation phase, an entangled pair of qubits is created. These two entangled qubits, denoted as $\mathit{EPR}_{tx}$ and $\mathit{EPR}_{rx}$, are then distributed via quantum channels to the respective nodes where the source and destination qubits reside. During the pre-processing phase, the qubit to be transmitted ($q_{tx}$) and $\mathit{EPR}_{tx}$ undergo a quantum measurement, producing two bits of classical information. These two bits are then transmitted to the destination node through a classical communication channel. Finally, in the post-processing phase, the received classical bits are used to determine which of four possible quantum operations should be applied to $\mathit{EPR}_{rx}$. This operation ensures that the quantum state of $q_{tx}$ is faithfully transferred to $\mathit{EPR}_{rx}$, completing the teleportation process.

\section{Reference Architecture}
\label{sec:reference_arch}

We consider a reference architecture representative of a fully cryogenically-controlled multi-core quantum system. This architectural model captures the expected evolution of scalable quantum platforms, where both quantum cores and their control logic are integrated at cryogenic temperatures to reduce latency and preserve coherence. In this section, we provide a description of the architectural components, including the organization of QCs, the communication network, and the classical infrastructure required for synchronization and control. This reference model serves as the foundation for evaluating how different architectural and micro-architectural parameters impact execution time in the subsequent sections.

\begin{figure}
    \centering
    \includegraphics[width=0.9\columnwidth]{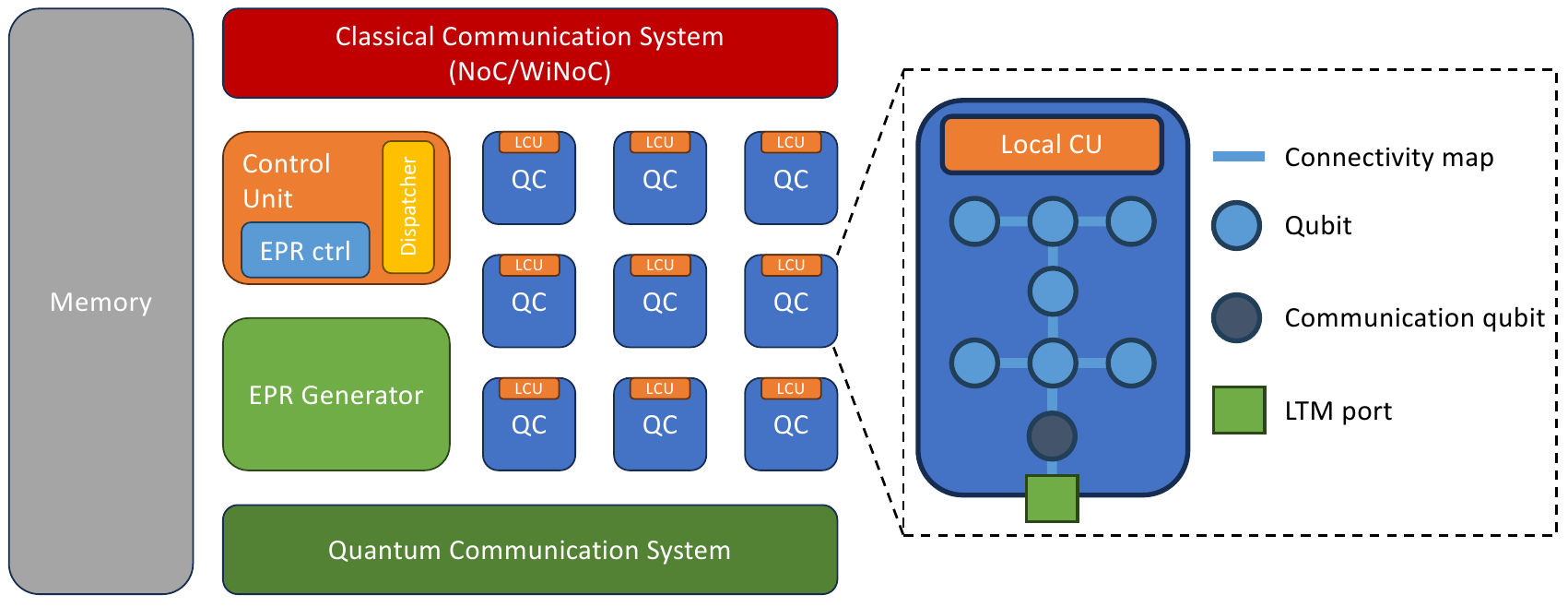}
    \caption{Main modules of the proposed architecture.}
    \label{fig:modules_and_qcore}
\end{figure}

Figure~\ref{fig:modules_and_qcore} illustrates the key modules that define the reference architecture, which will be used throughout the remainder of this paper. The \emph{Memory} module stores the program instructions to be executed (Sec.~\ref{ssec:circuit2assembly}). The \emph{Control Unit} fetches these instructions from memory, decodes them, and dispatches them to the quantum cores for execution (Sec.~\ref{ssec:control_unit}). The \emph{EPR Generator} creates EPR pairs between quantum cores to support the teleportation protocol (Sec.~\ref{ssec:epr_generator}). The array of \emph{Quantum Cores} executes the quantum gates (Sec.~\ref{ssec:qcore}). The \emph{Quantum Communication System} functions as the communication backbone for qubit transfers. Meanwhile, 
the \emph{Classical Communication System} manages classical communication between the various modules and plays a central role in orchestrating the entire system.

Alternative approaches such as direct quantum state transfer~\cite{bose_prl03} or remote gate execution~\cite{vanmeterqic06} have also been proposed for inter-core communication. However, these methods often impose stricter requirements on qubit coherence, channel losses, and direct connectivity, which makes them less practical in the near term. Teleportation, by contrast, separates the entanglement distribution stage from data transmission and integrates naturally with modular architectures, which is why we adopt it as the baseline communication protocol in \qcomm{}.

\subsection{Quantum Core}
\label{ssec:qcore}
A quantum core (see the right side of Figure~\ref{fig:modules_and_qcore}) consists of a set of physical qubits, one or more light-to-matter (LTM) ports, and a local control unit. An LTM port is a physical interface that enables a quantum core to interact with remote qubits by entangling its matter qubits with photons used as flying qubits in optical communication (see Sec.~\ref{ssec:epr_generator} for more details). Physical qubits connected to LTM ports are referred to as \emph{communication qubits}. Gates can only be applied between qubits that are directly connected in accordance with a connectivity map. The local control unit decodes the instructions received from the global control unit (Sec.~\ref{ssec:control_unit}) and generates the corresponding control signals to execute the instructions (i.e., gates) within its QC.
%\begin{figure}
%    \centering
%    \includegraphics[width=0.5\columnwidth]{figs/qcore.pdf}
%    \caption{A quantum core with seven physical qubits, one communication qubit and one LTM port.}
%    \label{fig:qcore}
%\end{figure}

\begin{figure}
    \centering
    \includegraphics[width=0.5\columnwidth]{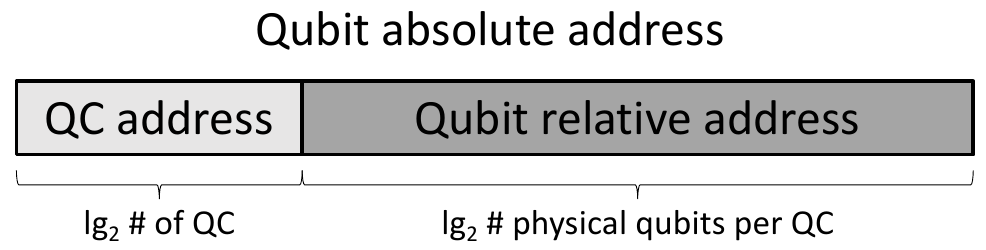}
    \caption{The address of a physical qubit is partitioned into the QC address and the local qubit address.}
    \label{fig:qubit_address}
\end{figure}

In a system with $M$ QCs, each containing $Q$ physical qubits, addressing a physical qubit requires $\lceil \lg_2 (M \times Q) \rceil$ bits. The qubit absolute address is thus divided into two parts: the QC address and the relative address of the physical qubit, as illustrated in Figure~\ref{fig:qubit_address}.

\subsection{From Circuit to Assembly}
\label{ssec:circuit2assembly}
\begin{figure}
    \centering
    \includegraphics[width=0.99\columnwidth]{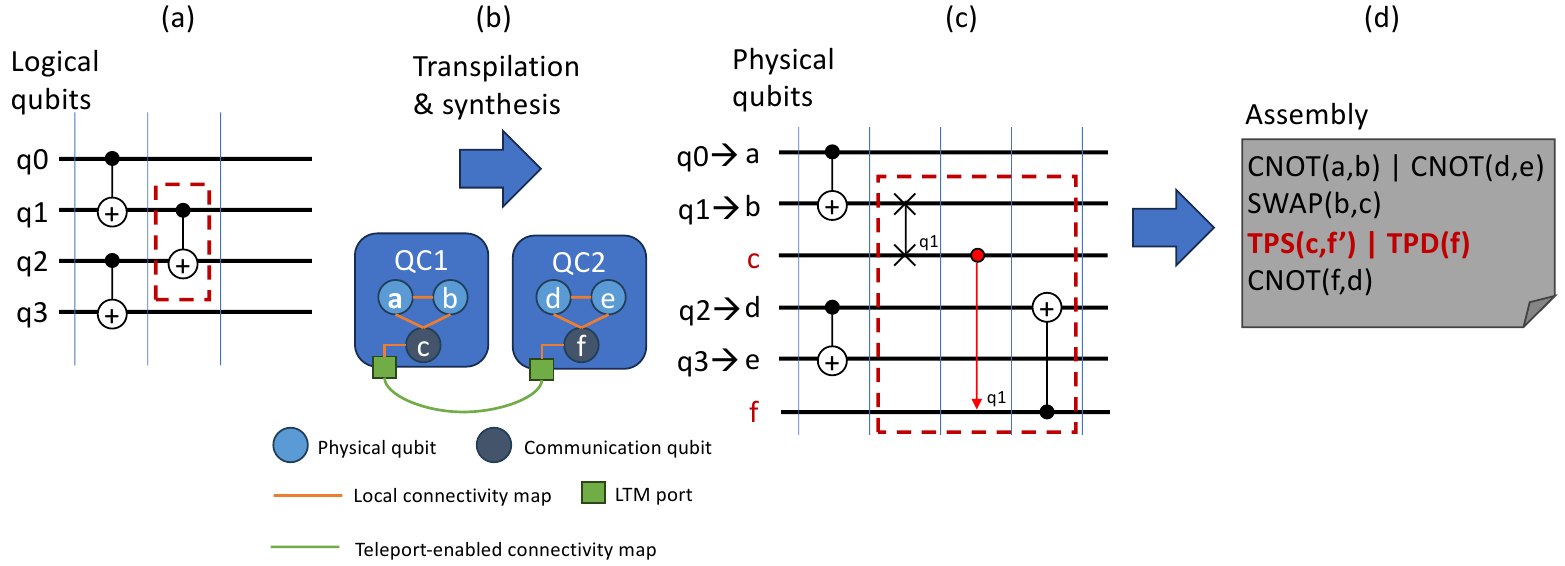}
    \caption{From circuit to assembly code. Logical circuit (a). Compilation phase (b). Synthesized circuit (c). Assembly code (b).}
    \label{fig:circuit}
\end{figure}

Consider the logic circuit (using logical qubits) shown in Figure~\ref{fig:circuit}a. The circuit is compiled by taking into account the total number of physical qubits, available gates, and the \emph{global connectivity map}. The global connectivity map refers to the combination of the local connectivity maps within each QC and the \emph{teleport-enabled connectivity map}, as illustrated in Figure~\ref{fig:circuit}b. The teleport-enabled connectivity map is formed by considering that communication qubits in $\textrm{QC}_i$ are connected to communication qubits in $\textrm{QC}_j$, where $i \ne j$.

The result of the compilation is a circuit where logical qubits are mapped onto physical qubits, and only the gates supported by the QCs are used. For the given example, the compiled circuit is shown in Figure~\ref{fig:circuit}c. Logical qubits $q0$, $q1$, $q2$, and $q3$ are mapped to physical qubits $a$, $b$, $d$, and $e$, respectively. The vertical lines divide the circuit into slices, which represent sets of gates that can be executed in parallel, as they involve disjoint sets of qubits. The first two CNOT gates can be executed in parallel on QC1 and QC2. However, the CNOT between logical qubits $q1$ and $q2$ must be executed between physical qubits $b$ and $d$, which are mapped to different QCs. Therefore, it requires teleporting either $b$ to QC2 or $d$ to QC1. Suppose the compiler decides to teleport $b$ to QC2. To achieve this, $b$ is first swapped with $c$, and then the teleportation protocol is applied to transfer the state of $c$ to $f$. This operation is represented by a red arrow from the source qubit to the destination qubit. Now, the CNOT can be executed between $f$ and $d$.

The graphical representation of the circuit involving physical qubits is translated into a textual format, specifically assembly code, as shown in Figure~\ref{fig:circuit}d. In the assembly code, each line corresponds to an \emph{instruction bundle}, which represents a slice of the compiled circuit. An instruction bundle consists of a set of instructions that can be executed in parallel. Such a bundle-based control model is inspired by the Very Long Instruction Word (VLIW) paradigm~\cite{fisher_isca83}. This design choice is motivated by several key considerations. First, quantum processors typically support a limited set of native gates, resulting in relatively simple instruction formats and reduced control flow complexity. Second, the mapping between logical and physical qubits, as well as the routing required to satisfy both local and teleportation-enabled connectivity constraints, can be effectively handled at compile time. This allows shifting complexity away from the hardware and into the compiler, reducing the need for dynamic scheduling and instruction decoding logic within each quantum core. Finally, minimizing hardware complexity is particularly important in the context of cryogenically-controlled architectures, where reducing the footprint and energy consumption of classical control electronics is critical to system scalability and thermal management.

In addition to the instructions for implementing quantum gates, two specific instructions, \texttt{TPS} and \texttt{TPD}, are introduced to execute the teleportation protocol. The instruction \texttt{TPS(c,f')} implements the first part of the teleportation protocol, executed by the source QC (which hosts $c$), and consists of the following phases: 
\begin{enumerate} 
    \item EPR pair generation. 
    
    \item EPR distribution, meaning the transmission of the entangled pairs through quantum channels to the LTM ports connected to the communication qubits $c$ and $f$. In all instructions, the operand refers to the qubit's relative address. However, in \texttt{TPS(c,f')}, the qubit $f$ is marked with a tick symbol to indicate its absolute address. This is necessary because the source QC must know the destination QC's address to send the two-bit classical information required for implementing the teleportation protocol.
    
    \item Pre-processing, where the qubit to be transmitted ($c$) and one of the two entangled qubits (delivered to the LTM port connected to $c$) are pre-processed and measured. 
    
    \item Classical communication, in which the two bits of classical information generated in the previous phase are transmitted to the destination QC (which hosts $f$). 
\end{enumerate}
The instruction \texttt{TPD(f)} implements the second part of the teleportation protocol, carried out by the destination QC (which hosts $f$). This part begins when the destination receives the two bits of classical information from the source QC. These bits are used to determine one of four quantum operations to apply to the entangled qubit (delivered to the LTM port connected to $f$), which is then swapped with $f$. This completes the transfer of the state of $c$ to $f$.

\subsection{Bundle Format}
\label{ssec:bundle_format}
\begin{figure}
    \centering
    \includegraphics[width=0.99\columnwidth]{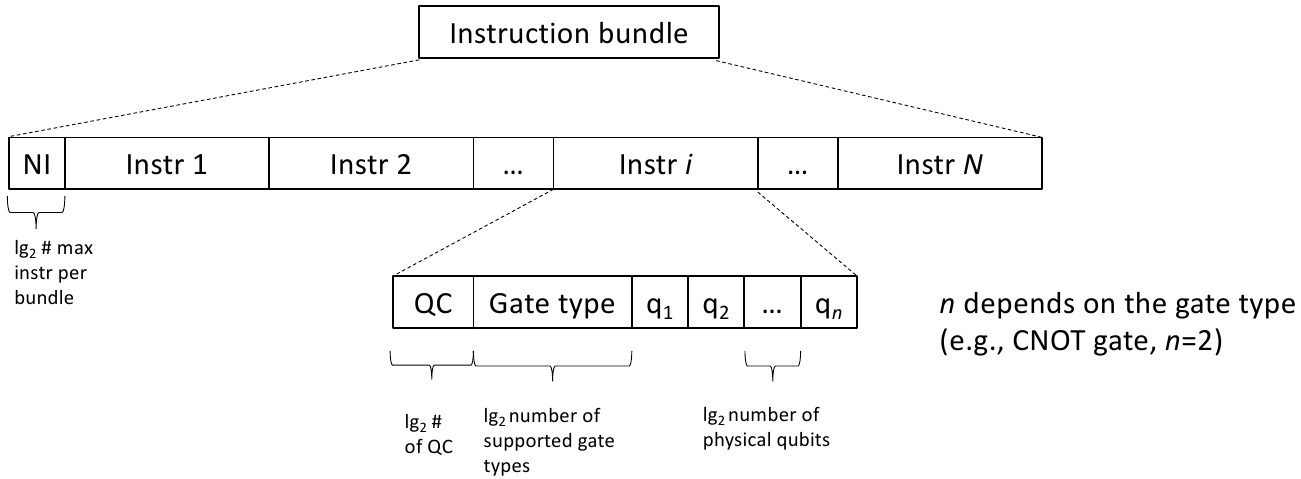}
    \caption{Format of the instruction bundle.}
    \label{fig:bundle}
\end{figure}
We refer to the instruction bundle format shown in Figure~\ref{fig:bundle}. The length of the instruction bundle is variable, meaning it can contain a different number of instructions. The number of instructions in the bundle is encoded in the first few bits. Assuming a maximum of $NI$ instructions per bundle, this requires $\lceil \lg_2 NI \rceil$ bits. The instructions within the bundle are concatenated sequentially.

The format of an individual instruction is as follows. For a system with $M$ QCs, the first $\lceil \lg_2 M \rceil$ bits encode the QC address. The next set of bits encodes the gate type. If the system supports $G$ different gate types, $\lceil \lg_2 G \rceil$ bits will be used to encode the gate type. The number of qubits involved in the instruction varies depending on the gate type. Assuming a homogeneous system where each QC has the same number, $Q$, of physical qubits, the qubit address in a QC requires $\lceil \lg_2 Q \rceil$ bits. Thus, the size in bits of a bundle $B$ is:
\begin{equation}
    BS(B) = \lceil \lg_2 NI \rceil + \sum_{i=1}^{N(B)}\left( \lceil \lg_2 M \rceil + \lceil \lg_2 G \rceil + O(B(i)) \lceil \lg_2 Q \rceil \right)
    \label{eqn:bundle_size}
\end{equation}
where $N(B)$ denotes the number of instructions in $B$, $O(I)$ indicates the number of operands for instruction $I$, and $B(i)$ represents the $i$-th instruction in bundle $B$. 

We note that when an instruction (i.e., a gate) is executed within a QC (whose address is encoded in the first bits of the instruction), the qubits specified in the instruction correspond to the physical qubits of that core. Therefore, only their relative addresses are required. The sole exception is the instruction \texttt{TPS(s,d')}, where $d'$ refers to the absolute address of the destination qubit $d$. This is because implementing the teleportation protocol requires sending two bits of classical information from the current QC to the destination QC, whose address is derived from the most significant bits of the absolute address of the destination qubit.

\subsection{Control Unit}
\label{ssec:control_unit}
Similar to classical computer architectures, the role of the control unit (CU) is to fetch an instruction bundle from memory, decode it, and generate the appropriate control signals to execute the instructions within the bundle in their respective QCs. The CU consists of two main modules: the \emph{dispatcher} and the \emph{EPR control}. 

\begin{comment}
%% previous version
    The dispatcher's role is to disaggregate the instruction bundle into individual instructions and deliver them to the appropriate QC. The target core is identified based on the address of the qubits involved in the instruction. As described in Sec.~\ref{ssec:qcore} and illustrated in Figure~\ref{fig:qubit_address}, the most significant bits of the qubit address specify the QC. Once the CU identifies the target QC, it modifies the instruction by replacing the qubit addresses with the corresponding local qubit addresses within that QC. This approach reduces the amount of information transmitted from the CU to the QC array, thereby decreasing classical communication traffic within the system.
\end{comment}

The dispatcher's role is to disaggregate the instruction bundle into individual instructions and deliver them to the appropriate QC. To reduce NoC traffic, the dispatcher translates global (absolute) qubit addresses into local addresses before dispatching the instruction to the target QC. This conversion is performed on-the-fly and requires no lookup tables or additional latency. In our model of a homogeneous system, where each QC contains the same number of qubits $Q$, the dispatcher derives the local address simply by masking the first $\log_2 M$ bits of the global address (with $M$ being the number of QCs) and retaining the remaining $\log_2 Q$ bits (see Figure~\ref{fig:qubit_address}). Due to the simplicity of this bit-masking operation, no timing or energy overhead is expected, even in a cryogenic control environment.

The EPR control module is responsible for generating the control signals that configure the EPR generator module to produce the EPR pairs required to support teleportation when the \texttt{TPS} instruction is executed.

The logic of the CU is simple enough to be implemented as a finite state machine, which is detailed in Appendix~\ref{apdx:cu_fsm}.

\subsection{EPR Generator}
\label{ssec:epr_generator}

% The functionality of entanglement generation and distribution is a fundamental component of quantum teleportation~\cite{cacciapuoti_tcomm20}. 
% Entanglement typically arises when two (or more) quantum systems interact or are jointly created in such a way that their quantum states become intrinsically correlated. 
% In distributed architectures, where the source and destination QCs are located at separate nodes, one part of the entangled pair must be transferred from its point of creation to the remote node, requiring a reliable quantum distribution mechanism.

\begin{figure}
\centering
    \begin{minipage}{0.8\textwidth}
    \centering
    \includegraphics[width=0.9\textwidth]{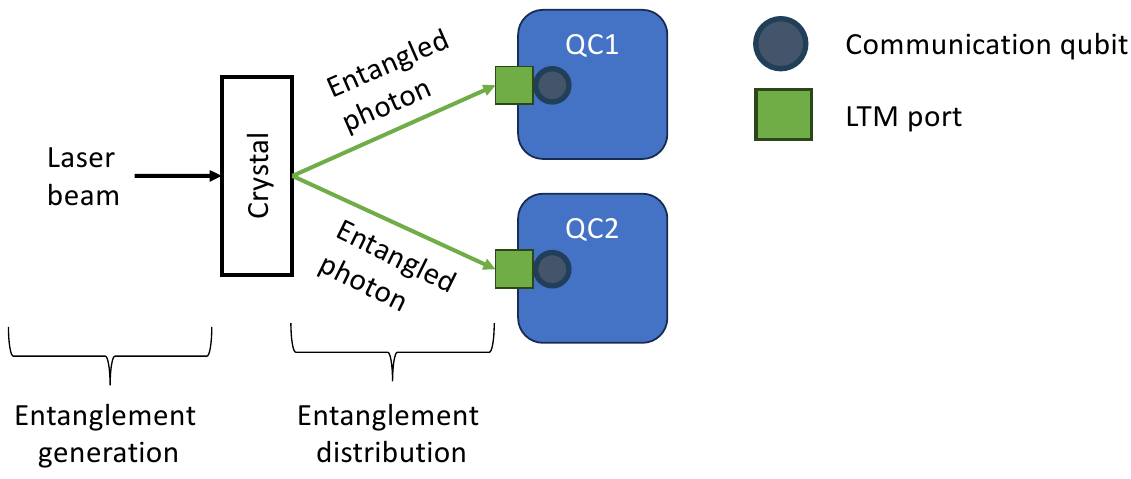}
    \subcaption{EPR generation at the \emph{mid-point}: A laser beam is directed at a non-linear crystal, which occasionally produces pairs of polarization-entangled photons by splitting an incoming photon beam.}\label{fig:epr_mid}
    \end{minipage}%
    \\
    \begin{minipage}{0.8\textwidth}
    \centering
    \includegraphics[width=0.9\textwidth]{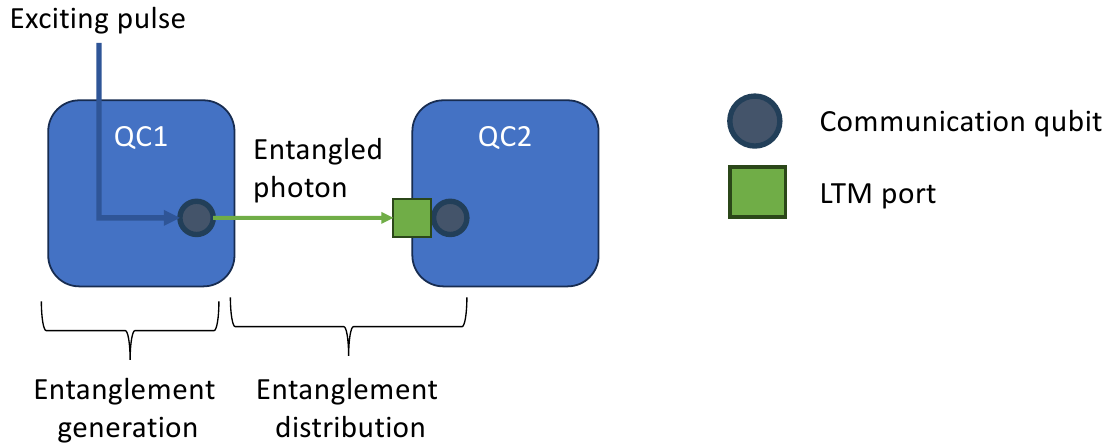}
    \subcaption{EPR generation at \emph{source}: An atom, strongly coupled to an optical cavity, is excited by a laser beam. The emitted photon escapes from the cavity, travels as a wave packet through a cable, and enters an optical cavity at a second node, establishing entanglement between the two remote atoms.}\label{fig:epr_source}
    \end{minipage}
    \\
    \begin{minipage}{0.8\textwidth}
    \centering
    \includegraphics[width=0.9\textwidth]{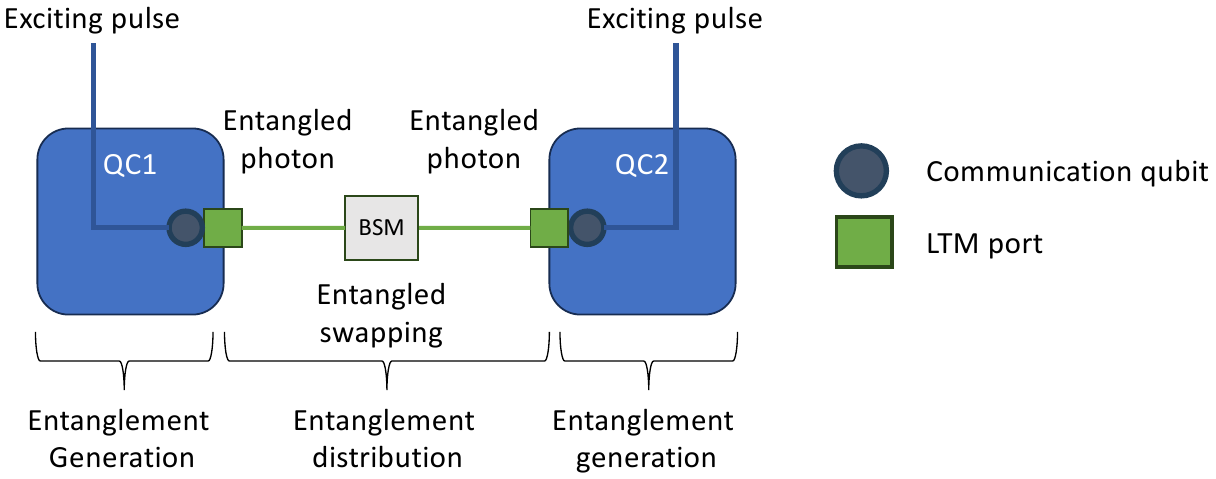}
    \subcaption{EPR generation at \emph{both end-points}: Two atoms, each confined in an optical cavity, are simultaneously excited by a laser pulse. This excitation results in the emission of two entangled photons. Upon measurement, a Bell-state measurement (BSM) projects the atoms into an entangled quantum state.}\label{fig:epr_both}
    \end{minipage}
\caption{Practical approaches for entanglement generation and distribution~\cite{cacciapuoti_tcomm20}. Regardless of where the entanglement is generated---whether at the mid-point (Figure~\ref{fig:epr_mid}), at the source (Figure~\ref{fig:epr_source}), or at both end-points (Figure~\ref{fig:epr_both})---a quantum link is essential to distribute the entanglement between QC1 and QC2.}\label{fig:epr_generation}
\end{figure}

In distributed architectures, where the source and destination QCs are located at separate nodes, one part of the entangled pair must be transferred from its point of creation to the remote node, requiring a reliable quantum distribution mechanism. A common and broadly adopted strategy in quantum networks employs \emph{flying qubits}, often realized as photons~\cite{northup_nature14}, to carry entanglement across physical distances. Regardless of the underlying hardware, the entanglement generation and distribution process can generally be classified into three functional paradigms~\cite{cacciapuoti_tcomm20}, each representing a different locus of entanglement creation and transfer as shown in Figure~\ref{fig:epr_generation}:
\begin{description}
    \item[EPR generation at the \emph{mid-point}] In this approach, entanglement is created at a central node between two remote QCs. A typical realization involves the use of a nonlinear process such as spontaneous parametric down-conversion, which produces entangled photon pairs that are then directed toward the two QCs. Upon arrival, each photon is converted from a flying qubit into a matter qubit using an interface such as a LTM transducer~\cite{afzelius_phytoday15}. This strategy is illustrated in Figure~\ref{fig:epr_mid} and is well-suited for architectures where entanglement sources are centrally accessible.

    \item[EPR generation at \emph{source}] Here, entanglement is initially established locally between a matter qubit and a photon at the source QC. The photon then travels through a quantum channel and interacts with a matter qubit at the destination QC, effectively transferring the entangled state between the two matter qubits. This method, often involving photon-mediated interactions with optical cavities, is shown in Figure~\ref{fig:epr_source} and allows for more distributed control of entanglement initiation~\cite{welte_phyreview18}.

    \item[EPR generation at \emph{both end-points}] A third approach involves exciting qubits at both source and destination nodes simultaneously, generating entangled photon-matter pairs locally. The emitted photons are then interfered at a beam-splitter, where a Bell State Measurement (BSM) probabilistically projects the two matter qubits into an entangled state~\cite{olmschenk_science09}. This scheme enables heralded entanglement generation and is compatible with a variety of physical qubit platforms, including Nitrogen-Vacancy (NV) centers in diamond~\cite{bernien_nature13} and superconducting transmons~\cite{kurpiers_nature18}, as illustrated in Figure~\ref{fig:epr_both}.
\end{description}

Across all these schemes, a key element is the transducer that mediates the interaction between stationary and flying qubits---regardless of whether this is implemented using optical cavities, waveguides, or microwave resonators. While the specific physical realizations differ, the architectural modeling adopted in this work abstracts these roles into functional modules: an entanglement generator, a quantum channel, and a matter-flying qubit interface.

In this work, the EPR generator is modeled as fully parallelized, meaning that multiple entangled pairs can be generated and distributed simultaneously across the available LTM ports. While this abstraction simplifies analysis and enables tractable performance evaluation, we acknowledge that it represents a non-trivial engineering assumption, particularly at large scales. Recent research indicates that the simultaneous generation of multiple EPR pairs is an active area of investigation~\cite{bravyi_arxiv24}, making this assumption relevant as a forward-looking design choice. Nevertheless, alternative models, such as time-multiplexed EPR generation, could be integrated into \qcomm{} in future versions to explore scalability trade-offs and the potential emergence of bottlenecks.

\subsection{Interconnect Architecture Options}
\begin{figure}
    \centering
    \includegraphics[width=0.9\columnwidth]{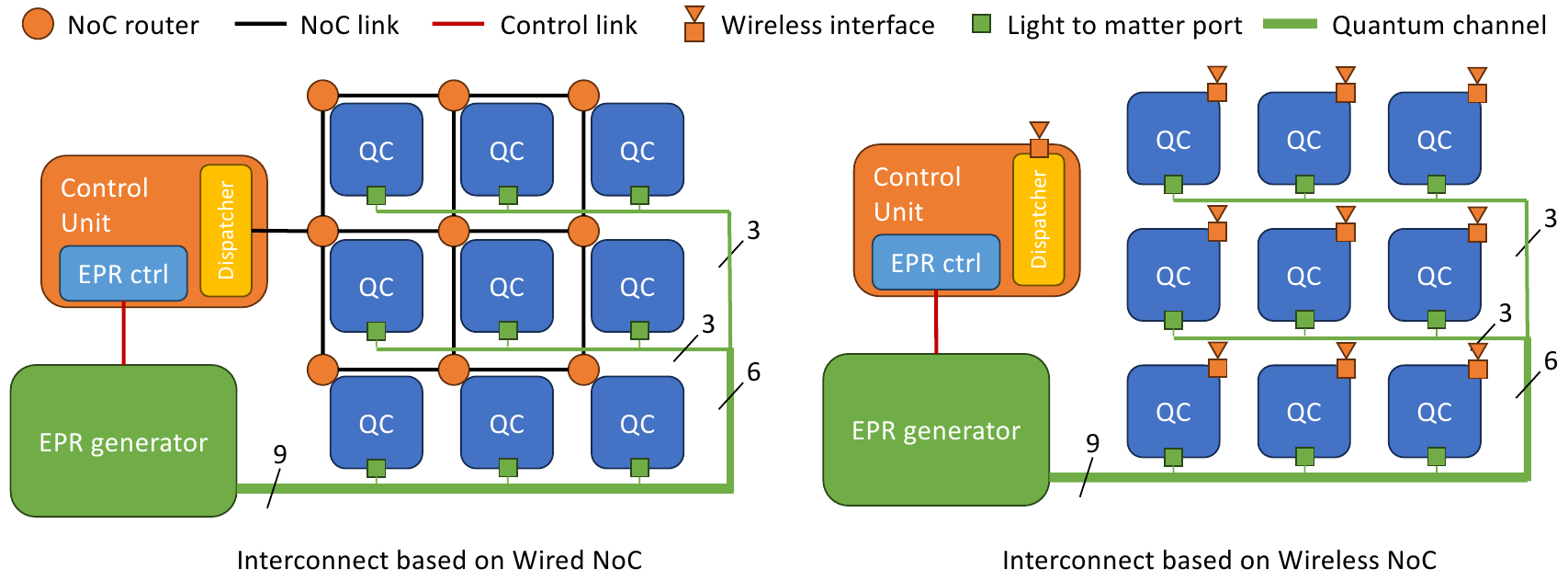}
    \caption{Interconnect option based on a wired NoC (left) and on a wireless NoC (right).}
    \label{fig:interconnect_options}
\end{figure}

We consider two interconnect options for communication between the CU and QCs, as well as between the QCs themselves as illustrated in Figure~\ref{fig:interconnect_options}. The first option is a traditional wired NoC~\cite{benini_computer02}. In this configuration, each QC is connected to a NoC router, and these routers are interconnected using a mesh topology. The CU, specifically the dispatcher, is also connected to the NoC. The EPR generator establishes point-to-point connections with the LTM ports in the QCs via dedicated quantum channels. If a QC has multiple LTM ports, each port is connected to the EPR generator through its own quantum channel.

The second option, replaces the wired NoC with a wireless NoC (WiNoC)~\cite{chiariotti_commsurv21}. In this case, each QC, as well as the dispatcher, is equipped with a wireless interface (WI), enabling wireless communication between the dispatcher and the QCs, and among the QCs themselves. The EPR generator remains connected to the QCs in the same manner as in the wired option.

\subsection{Execution Model Example}
\label{ssec:execution_model}
\begin{figure}
    \centering
    \includegraphics[width=0.8\columnwidth]{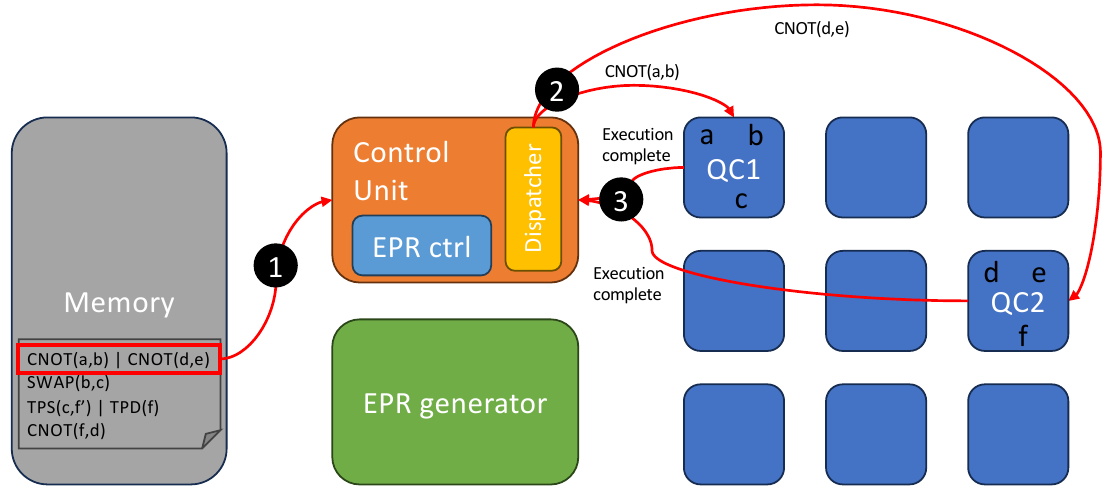}
    \caption{Phases involved in the execution of the local bundle $\langle$\texttt{CNOT(a,b) | CNOT(d,e)}$\rangle$.}
    \label{fig:cnotcnot}
\end{figure}

\begin{figure}
    \centering
    \includegraphics[width=0.8\columnwidth]{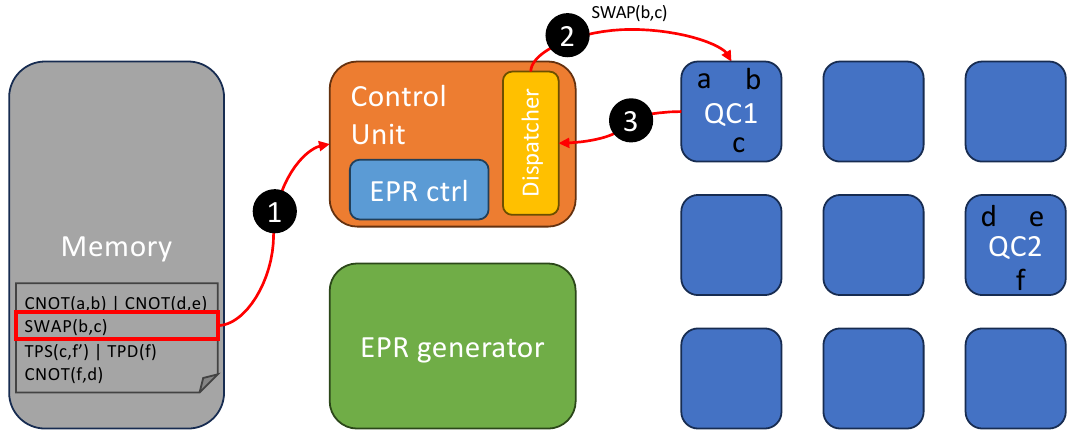}
    \caption{Phases involved in the execution of the local bundle $\langle$\texttt{SWAP(b,c)}$\rangle$.}
    \label{fig:swap}
\end{figure}

\begin{figure}
    \centering
    \includegraphics[width=0.8\columnwidth]{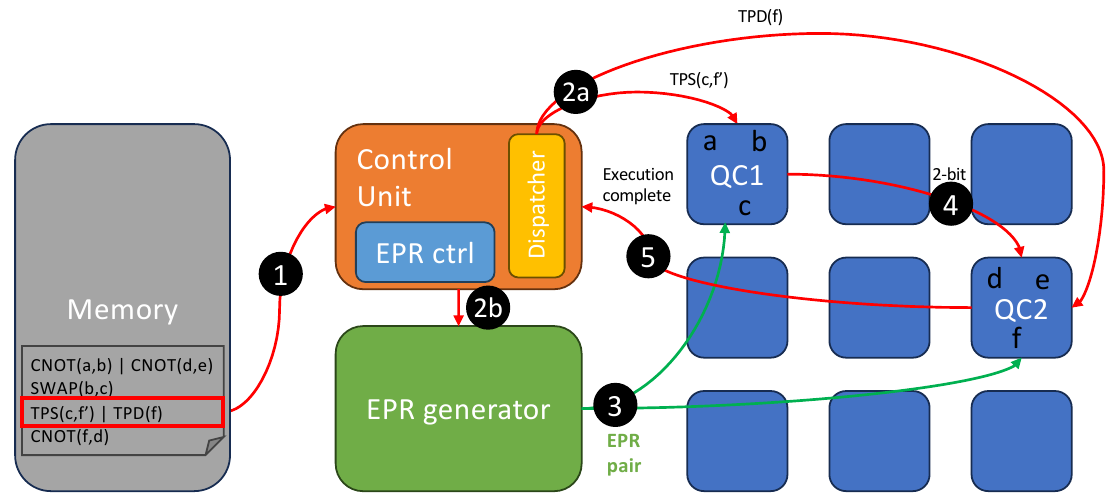}
    \caption{Phases involved in the execution of the remote bundle $\langle$\texttt{TPS(c,f) | TPD(f)}$\rangle$.}
    \label{fig:tpstpd}
\end{figure}

This section delineates the conceptual execution model of the proposed architecture. For the purpose of discussion, we refer to the same assembly code shown in Figure~\ref{fig:circuit}d. As three different approaches for entanglement generation and distribution are possible, in this section we refer to the EPR generation at the \emph{mid-point} and it will be assumed that: 1) the EPR Generator maintains point-to-point connectivity with any QC through quantum channels, and 2) it possesses the capability to concurrently generate multiple EPR pairs for any pair of QCs. Consequently, it is assumed that in a system comprising $M$ QCs, each with $L$ LTM ports, there exist $M \times L$ quantum channels linking the EPR Generator to individual LTM ports of each specific QC. The impact on the architecture and timing when the other entanglement generation and distribution approaches are considered are discussed in the Appendix~\ref{pdx:epr}. The different phases involved in the execution of a bundle are illustrated in Figures~\ref{fig:cnotcnot}–\ref{fig:tpstpd} and described below.

\paragraph{Execution of $\langle$\texttt{CNOT(a,b) | CNOT(d,e)}$\rangle$}
The CU fetches the instruction bundle $\langle$\texttt{CNOT(a,b) | CNOT(d,e)}$\rangle$ from memory (Figure~\ref{fig:cnotcnot}\cone). Since qubits $a$ and $b$ belong to QC1, and qubits $d$ and $e$ belong to QC2, the dispatcher sends the first \texttt{CNOT} to QC1 and the second to QC2 (Figure~\ref{fig:cnotcnot}\ctwo). The transmitted instructions are modified so that the qubit addresses refer to their local addresses. This is illustrated in the figure using the operator \texttt{L()}, which extracts the least significant bits of the qubit address, corresponding to the local qubit address (see the qubit address format in Figure~\ref{fig:qubit_address}). 

The local control units of QC1 and QC2 steer their respective QCs to execute the \texttt{CNOT} operations. Once the execution is complete, an \emph{execution complete} message is sent back to the CU to signal the completion of the instruction (Figure~\ref{fig:cnotcnot}\cthree). The CU applies a barrier over all issued instructions, which is only released when all execution complete messages associated with the issued instructions have been received. After that, the CU is ready to fetch the next instruction bundle.

\paragraph{Execution of $\langle$\texttt{SWAP(b,c)}$\rangle$}
The CU fetches the next instruction bundle from memory (Figure~\ref{fig:swap}\cone). This time, the bundle contains a single instruction, \texttt{SWAP(b,c)}, which is decoded and dispatched to QC1, as both $b$ and $c$ belong to QC1 (Figure~\ref{fig:swap}\ctwo). Once again, the dispatched instruction is modified by replacing the global addresses of the operand qubits with their local addresses. The instruction is then decoded by the local CU in QC1 and executed. After execution, the local CU sends an \emph{execution complete} message to the CU, indicating the instruction has finished (Figure~\ref{fig:swap}\cthree). The CU, which was waiting for a single execution completion message, is now ready to fetch the next instruction.

\paragraph{Execution of $\langle$\texttt{TPS(c,f') | TPD(f)}$\rangle$}
The CU fetches the instruction bundle $\langle$\texttt{TPS(c,f') | TPD(f)}$\rangle$, which implements the teleportation operation, aiming to teleport the quantum state of $c$ to $f$ (Figure~\ref{fig:tpstpd}\cone). The \texttt{TPS} instruction is dispatched to QC1, and the \texttt{TPD} instruction is sent to QC2 (Figure~\ref{fig:tpstpd}\ctwo a). This time, the address of the second operand in \texttt{TPS} is not replaced with its local address, as QC1 needs to know the address of $f$ in order to send the two classical bits of information from QC1 to QC2 during the first part of the teleportation protocol. Simultaneously, the EPR control module configures the EPR generator (Figure~\ref{fig:tpstpd}\ctwo b), which produces the EPR pair distributed to both QC1 and QC2 (Figure~\ref{fig:tpstpd}\cthree). The local CU of QC1 executes the pre-processing phase of the teleportation protocol and transmits the 2-bit classical information to QC2 (Figure~\ref{fig:tpstpd}\cfour). The CU of QC2 then performs the post-processing phase to complete the teleportation. Finally, it sends an \emph{execution complete} message to the CU (Figure~\ref{fig:tpstpd}\cfive), allowing it to fetch the next instruction.

\paragraph{Execution of $\langle$\texttt{CNOT(f,d)}$\rangle$}
The last instruction bundle is fetched, and the \texttt{CNOT} is executed in a manner similar to the \texttt{SWAP} instruction described earlier.

\section{Timing Model}
\label{sec:timing_model}
This section outlines the timing models used to estimate execution time. Based on the previously described execution model, the total execution time is the sum of the execution times of the instruction bundles that make up the assembly code. Let $P$ represent the program to be executed, and let $P(i)$ denote the $i$-th instruction bundle in $P$. The total execution time of $P$ is the sum of the execution times of all the instruction bundles in $P$:
\begin{equation}
  ExecutionTime(P) = \sum_{i=1}^{N(P)} ExecutionTime(P(i)) 
\end{equation}
where $N(P)$ represents the number of instruction bundles in $P$.

The execution time of a bundle depends on its composition. There are two main scenarios: one where the instruction bundle contains operations exclusively on local qubits, and another where it includes teleportation instructions. We refer to the first scenario as a \emph{local bundle} and the second as a \emph{remote bundle}.

\subsection{Timing Model for Local Bundle}
\begin{figure}
    \centering
    \includegraphics[width=0.99\columnwidth]{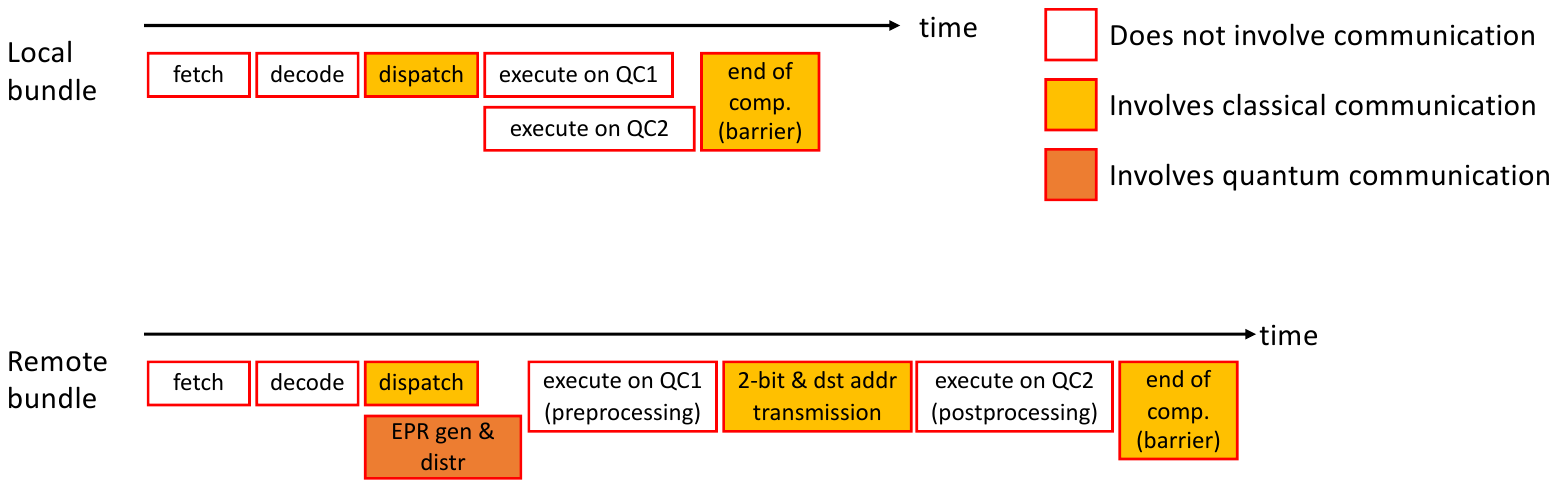}
    \caption{Execution timeline of a local bundle (top) and a remote bundle (bottom).}
    \label{fig:timeline}
\end{figure}

To compute the execution time of a local bundle, consider the example shown in Figure~\ref{fig:cnotcnot}. The top section of Figure~\ref{fig:timeline} illustrates the timing of the various execution phases described in Sec.~\ref{ssec:execution_model}, where all phases are executed sequentially. It also highlights the phases that involve both classical and quantum communication. While the actual execution of the instructions (gates) occurs in parallel, the total computation time is determined by the instruction with the longest execution time among all parallel instructions. Thus, the execution time of a local bundle $B$ is given by:
\begin{equation}
\begin{aligned}
ExecutionTime(B) &= FetchTime(B) + DecodeTime(B) + DispatchTime(B) + \\
                 &+ \max\{InstructionTime(B(i)), \ i=1,\ldots,N(B) \} + \\
                 &+ EndComputationTime(B)
\end{aligned}
\label{eqn:tl_local}
\end{equation}
where $B(i)$ is the $i$-th instruction, and $N(B)$ represents the number of instructions in bundle $B$.

\subsection{Timing Model for Remote Bundle}
To compute the execution time of a remote bundle, consider the example in Figure~\ref{fig:tpstpd}. The bottom part of Figure~\ref{fig:timeline} illustrates the timing of the different execution phases described in Sec.~\ref{ssec:execution_model}, with all phases executed sequentially. The dispatch, EPR generation, and distribution phases run in parallel, allowing the next phase, preprocessing, to begin once the slowest of the previous phases is completed. Thus, the execution time of a remote bundle $B$ is given by:
\begin{equation}
\begin{aligned}
ExecutionTime(B) &= FetchTime(B) + DecodeTime(B) + \\
                 &+ \max\{ DispatchTime(B), EPRTime(B) \} + \\
                 &+ PreProcTime(B) + ClasComTime(B) + PostProcTime(B) + \\
                 &+ EndComputationTime(B)
\end{aligned}
\label{eqn:tl_remote}
\end{equation}

\subsection{Detailed Breakdown of Timing Components}
The various terms that make up Eqs.~\eqref{eqn:tl_local} and~\eqref{eqn:tl_remote} are discussed below.

\paragraph{Fetch Time}
The fetch time refers to the time required to load an instruction bundle from main memory into the control unit. Considering the bundle format discussed in Sec.~\ref{ssec:bundle_format}, let $BS(B)$ denote the size of a bundle $B$ in bits, as defined in Eq.~\eqref{eqn:bundle_size}, and let $MB$ represent the memory bandwidth in bits per second. The fetch time for a bundle $B$ is then calculated as:
\begin{equation}
    FetchTime(B) = \frac{BS(B)}{MB}
\end{equation}

\paragraph{Decode Time}
We assume a linear model for the decode time as a function of the instruction bundle length, expressed as:
\begin{equation}
    DecodeTime(B) = d_1 + d_2 N(B)
    \label{eqn:decode_time}
\end{equation}
where $d_1$ represents the base decode time, and $d_2$ indicates the rate at which the decode time increases with the bundle length. 

\paragraph{Dispatch Time}
The dispatch time refers to the time required to distribute the instructions in the bundle to the QCs where the corresponding qubits are located. This time is influenced by the underlying communication infrastructure connecting the dispatcher to the QCs. For each instruction in the bundle, the dispatcher sends a packet to the relevant QCs, adjusting the instruction by converting the qubits’ absolute addresses to their local addresses. The total traffic volume generated by the dispatcher to send the instructions from bundle $B$ to QC $i$ is given by:
\begin{equation}
    TV(B,i) = \sum_{I \in I(B,i)} \left( \lceil \lg_2 G \rceil + O(I) \lceil \lg_2 Q \rceil \right)
\end{equation}
where $I(B,i)$ returns the set of instructions in $B$ that are designated for QC $i$.
Thus, the dispatch time can be calculated as:
\begin{equation}
    DispatchTime = \sum_{i=1}^{M} CCT(dispatcher, QC_i, TV(B,i)) 
\end{equation}
where $CCT(src, dst, vol)$ represents the classical communication time to send $vol$ bits from node $src$ to node $dst$ in the system. As mentioned earlier, this depends on the communication system being considered.

\paragraph{EPR Generation and Distribution Time}
We assume that the EPR generation time depends on the number of EPR pairs to be generated, as given by the following expression:
\begin{equation}
    EPRGenTime(B) = f_{epr}(\lvert QCT(B) \rvert / 2)
\end{equation}
where $QCT(B)$ represents the set of QCs involved in teleportation operations, if any, for the execution of the instructions in bundle $B$. The notation $\lvert \cdot \rvert$ indicates the cardinality of the set, and the division by two accounts for the fact that $f_{epr}$ takes as input the number of EPR pairs.

For the distribution time, we assume an ideal optical channel with no photon loss and a delay determined by the distance between the EPR generator and the destination node. This can be expressed as:
\begin{equation}
    EPRTraTime(B) = \max \{ dist(QC)/c', \ QC \in QCT(B) \}
\end{equation}
where $dist(QC)$ gives the distance of QC from the EPR generator, and $c'$ is the speed of light in the optical medium.

Therefore, the overall EPR generation and distribution time is calculated as:
\begin{equation}
    EPRGenTraTime(B) = EPRGenTime(B) + EPRTraTime(B)
\end{equation}
In our current model, we assume ideal optical channels with no photon loss, noise, or fidelity degradation, and we use a constant propagation delay based solely on the speed of light. This abstraction is adopted for tractability and to isolate the role of classical communication in execution time. Extending \qcomm{} to incorporate fidelity-aware teleportation and loss models is an important direction for future work.

\paragraph{Classical Communication Time}
The classical communication time, as referenced in Eq.~\eqref{eqn:tl_remote}, represents the time required to send the two bits of information necessary to complete the post-processing phase of the teleportation protocol. In addition to these two bits, the message sent from the source QC to the destination QC will include the address of the destination qubit. Therefore, the classical communication time accounts for the time taken to transmit such messages, which naturally depends on the underlying communication system. It is calculated as:
\begin{equation}
    ClasComTime(B) = \sum_{(src,dst) \in QCT(B)} CCT(src, dst, 2+\lceil \lg_2 (M \cdot Q) \rceil) 
\end{equation}

\section{\qcomm{} Simulator}
\label{sec:qcomm}
\begin{figure}
    \centering
    \includegraphics[width=0.6\columnwidth]{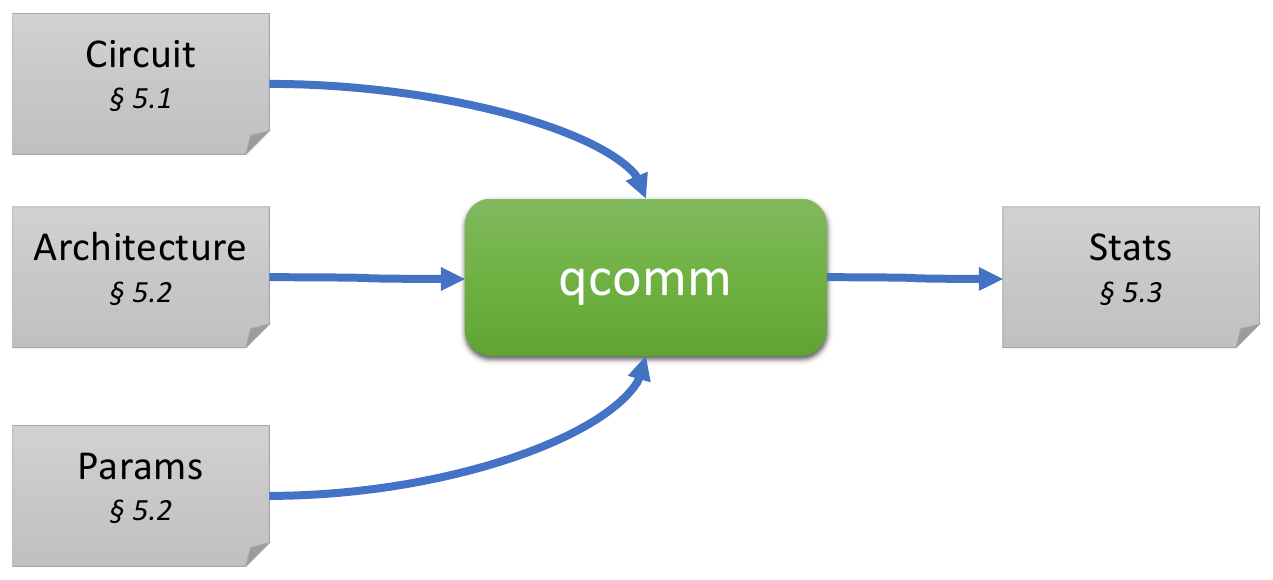}
    \caption{Inputs and output of \qcomm.}
    \label{fig:qcomm-io}
\end{figure}

We developed an open-source simulator, called \qcomm~\cite{palesi_mcsoc24}. The simulator implements the execution model described in Section~\ref{ssec:execution_model} and the timing model described in Section~\ref{sec:timing_model}. \qcomm{} gets in input the description of the quantum circuit, the description of the architecture, and the physical parameters. It produces in output the execution statistics (Fig.~\ref{fig:qcomm-io}). Each of the above elements, is presented in the following.

\subsection{Circuit Representation}
\begin{comment}
\begin{figure}
    \centering
    \includegraphics[width=0.6\textwidth]{figs/qcomm-circuit.pdf}
    \caption{Example of a circuit representation.}
    \label{fig:qcomm_circuit}
\end{figure}
A quantum circuit is described as a sequence of time slices. Each slice defines a set of gates that can be executed concurrently. Gates within a slice can be executed simultaneously if they share no input qubits. An example circuit description is provided in Figure~\ref{fig:qcomm_circuit}. Each line in the circuit file represents a slice, encoded as a list of gates. An $n$-input gate is represented by a tuple containing the qubit indices. For example, in Figure~\ref{fig:qcomm_circuit}, the second line represents the second slice. This slice includes a single-input gate involving qubit $q_2$ and a three-input gate involving qubits $q_3$, $q_7$, and $q_8$.
\end{comment}
A quantum circuit is described as a sequence of time slices. Each slice defines a set of quantum gates that can be executed concurrently. Gates within a slice can be executed simultaneously if they act on disjoint sets of qubits. The format used by \qcomm{} encodes each slice as a list of gates, where each gate is represented as a tuple of qubit indices indicating the qubits it operates on. For instance, a two-input gate like a CNOT is represented by a tuple of two qubit indices, while a three-qubit gate (e.g., Toffoli) would be a tuple of three indices.

% To facilitate usability, the \qcomm{} distribution includes a utility script that converts OpenQASM~\cite{cross_qst22} circuit descriptions into the expected slice-based format. This allows users to import circuits generated or transpiled using common quantum programming frameworks and simulate their behavior within the \qcomm{} environment.

In the current distribution of \qcomm{}, we provide a companion tool named \texttt{qasm2qcomm}. This command-line utility parses OpenQASM~2.0~\cite{cross_qst22} circuits (via Qiskit), inlines user-defined gates, decomposes them into native operations, removes measurement statements, and outputs a dependency-respecting sequence of parallel gate slices compatible with the simulator. This enables \qcomm{} to directly process circuits generated by existing compiler toolchains such as Qiskit, t$\mid$ket$\rangle$, or Cirq. At present, qubit-to-core assignment must be specified manually by the user; however, future versions of \qcomm{} will support compiler-driven mappings, thereby facilitating tighter integration between quantum compilers and architectural-level simulation.

\begin{table}
  \caption{Architectural parameters}
  \label{tab:arch_params}
  \begin{tabularx}{\textwidth}{lX}
    \toprule
    Parameter name &  Description \\
    \midrule
    \texttt{mesh\_x} and \texttt{mesh\_y} & Number of QCs in the horizontal and vertical dimensions, respectively. The total number of QCs is the product of \texttt{mesh\_x} and \texttt{mesh\_y}.\\

    \texttt{link\_width} & Width (number of data lines) of the NoC link. It also determines the flit size (a unit of data transferred over the NoC). A communication packet of $n$ bits is divided into $n/$\texttt{link\_width} flits, each with a size of \texttt{link\_width} bits.\\

    \texttt{qubits\_per\_core} & Total number of physical qubits per QC. Thus, the total number of physical qubits in the system is \texttt{mesh\_x} $\times$ \texttt{mesh\_y} $\times$ \texttt{qubits\_per\_core}.\\

    \texttt{ltm\_ports} & Number of LTM ports per QC. It determines the maximum number of concurrent teleportations that can involve the same QC.\\
    
    \texttt{wireless\_enabled} & Whether the communication system is implemented by a NoC or a wireless NoC (WiNoC)~\cite{deb_jestcs12}.\\

    \texttt{radio\_channels} & Number of radio channels available in the WiNoC (if enabled by \texttt{wireless\_enabled}).\\

    \texttt{teleportation\_type} & Selects between \emph{all-to-all} teleportation and \emph{split} teleportation. \\ 

    \texttt{dst\_selection\_mode} & Defines the policy for selecting the destination QC when a multi-input gate involves qubits from different QCs. \\ % Two policies are currently implemented: \emph{load-aware} and \emph{load-independent}. In the \emph{load-aware} policy, the QC with the fewest allocated logical qubits (i.e., the most available physical qubits) is chosen as the destination. In the \emph{load-independent} policy, the destination QC is determined based on the mapping of the qubit used for the last input of the gate. \\
    \bottomrule
  \end{tabularx}
\end{table}

\begin{table}
  \caption{Physical parameters and micro-architectural parameters.}
  \label{tab:phy_params}
  \begin{tabularx}{\textwidth}{lX}
    \toprule
    Parameter name &  Description \\
    \midrule
    \texttt{gate\_delays} &  A mapping $\langle \texttt{gate\_name}, \texttt{delay} \rangle$ specifying the execution delay of each quantum gate type. \\
    
    \texttt{epr\_delay} & Mean of the EPR pair generation time.\\

    \texttt{dist\_delay} & EPR pair distribution time.\\

    \texttt{pre\_delay} & Delay of the pre-processing block for teleportation.\\

    \texttt{post\_delay} & Delay of the post-processing block for teleportation.\\

    \texttt{noc\_clock\_time} & Clock time of the NoC which determines the hop time. The latter includes both the router delay and the link delay.\\
    
    \texttt{wbit\_rate} & WiNoC bit-rate used only if the \texttt{wireless\_enabled} flag in the architecture file is set.\\

    \texttt{token\_pass\_time} & When the communication system is a WiNoC, the medium access control mechanism is based on token passing~\cite{deb_jestcs12}, where only the wireless interface (WI) holding the token can transmit. This parameter represents the time spent by the token to be moved from one WI to another. \\

    \texttt{memory\_bandwidth} & Memory bandwidth used with \texttt{bits\_instruction} to compute the time spent for fetching the instruction bundles from the memory.\\

    \texttt{bits\_instruction} & Number of bits used for encoding an instruction in the bundle. It depends by the number of instructions (i.e., gates) in the instruction set.\\

    \texttt{decode\_time} & Time for decoding an instruction.\\
    \bottomrule
  \end{tabularx}
\end{table}

\subsection{Architectural and Physical Parameters}
The architecture file defines the resources of the QCs and the characteristics of the NoC, describing how these QCs are interconnected. These specifications are provided through a set of architectural parameters, as detailed in Table~\ref{tab:arch_params}. These parameters include the system size, expressed as the number of QCs, determined by the product of \texttt{mesh\_x} and \texttt{mesh\_y}. They also cover NoC/WiNoC-related parameters such as link width, number of radio channels, and quantum-specific architectural parameters like the number of qubits per core and the number of LTM ports. Additional architectural parameters include the teleportation type.

Currently, two teleportation approaches are supported, controlled by the \texttt{teleportation\_type} option, reflecting the EPR generation and distribution methods discussed in Section~\ref{ssec:epr_generator} and illustrated in Figure~\ref{fig:epr_generation}. In \emph{all-to-all} teleportation, the EPR generator is assumed to have point-to-point connections with all QCs, enabling the simultaneous generation of multiple EPR entangled pairs. This allows single-hop teleportation between any pair of QCs. In \emph{split} teleportation (also referred to as multi-hop teleportation), EPR entangled pairs are generated only between directly connected QCs. Therefore, if teleportation involves two QCs that are not directly connected, multiple teleportation steps are performed between intermediate QCs along a path linking the source and destination QCs. In the current implementation, a mesh topology is used, and XY routing determines the path along which teleportation is split.

\begin{comment}
    Currently, two teleportation approaches are supported, controlled by the \texttt{teleportation\_type} option, reflecting the EPR generation and distribution methods discussed in Sec.~\ref{ssec:epr_generator}. In \emph{all-to-all} teleportation, the EPR generator is assumed to have point-to-point connections with all QCs, enabling the simultaneous generation of multiple EPR entangled pairs. This allows single-hop teleportation between any pair of QCs. In \emph{split} teleportation, EPR entangled pairs are generated only between directly connected QCs. Therefore, if teleportation involves two QCs that are not directly connected, multiple teleportation steps are performed between intermediate QCs along a path linking the source and destination QCs. In the current implementation, a mesh topology is used, and XY routing determines the path along which teleportation is split.
\end{comment}

Another architectural parameter, \texttt{dst\_selection\_mode}, defines the policy for selecting the destination QC during teleportation. When executing a two-input gate involving qubits located in different QCs, a decision must be made on whether to teleport the first qubit to the QC of the second qubit or vice versa. In \emph{load-aware} mode, the QC with more available qubits is selected. In \emph{load-independent} mode, the QC where the second qubit resides is chosen.

The parameters file specifies the physical and micro-architectural parameters used in the simulation. These parameters, listed in Table~\ref{tab:phy_params}, include gate delay as well as delay and data rate metrics for the classical communication system (NoC and WiNoC).

The parameter \texttt{gate\_delays} specifies gate latencies as a mapping between gate names and their corresponding execution delays, i.e., $\langle \texttt{gate\_name}, \texttt{delay} \rangle$. The \texttt{noc\_clock\_time} sets the NoC clock frequency, which is used to calculate the delay for a flit to traverse a router and a link.

The \texttt{wbit\_rate} and \texttt{token\_pass\_time} parameters pertain to the WiNoC. Specifically, \texttt{wbit\_rate} represents the wireless transmission bit rate per radio channel, while \texttt{token\_pass\_time} defines the time required for the token to pass from one wireless interface (WI) to the next. The current version of the simulator employs a classical token-based medium access control (MAC) mechanism, where only the node holding the token is allowed to transmit. If multiple radio channels are available, multiple tokens are used.

The \texttt{memory\_bandwidth} parameter, together with \texttt{bits\_instruction} (which specifies the number of bits used to encode an instruction), determines the instruction fetch time. Finally, the \texttt{decode\_time} parameter defines the time required to decode an instruction within an instruction bundle.

\subsection{Statistics}
The simulation generates a report containing various execution statistics, as detailed in Table~\ref{tab:stats}. Key metrics include the total number of inter-core communications, as well as the average and peak throughput of classical communication. This encompasses NoC/WiNoC traffic required for teleportation, instruction dispatch, and other control messages.

The report also provides a breakdown of communication time, detailing its individual components. When the \emph{detailed} option is specified in the command line, the report is extended to include a more granular analysis. This includes detailed inter-core communication statistics for each QC pair, as well as, for each qubit, the number of operations performed and the number of teleportations it undergoes.

\begin{table}
  \caption{Statistics provided by \qcomm.}
  \label{tab:stats}
  \begin{tabularx}{\textwidth}{lX}
    \toprule
    Figure & Description \\
    \midrule
    Executed gates & Number of gates in the circuit that have been executed.\\

    Inter-core comms & Total number of inter-core communications.\\

    Inter-core traffic & Total number of qubits transferred across QCs.\\

    Inter-core comm-map & Number of inter-core communications between any pair of QCs. \\
    
    Throughput & Peak and average rate of classical communication observed during the execution of the circuit. This refers to the communication overhead (either on the NoC or WiNoC) required to implement the teleportation protocol.\\
    
    Core utilization & Average, minimum, and maximum number of qubits in a QC. Initially, the logical qubits of the circuit are uniformly mapped to the QCs. In other words, in a circuit with $n$ qubits, each QC will host $m=n$/(\texttt{mesh\_x} $\times$ \texttt{mesh\_y}) qubits, where $m$ must be less than or equal to \texttt{qubits\_per\_core}. As the simulation progresses, some qubits are redistributed from one QC to another, thus varying the distribution of qubits across the QCs.\\

    Communication time & Portion of the execution time spent on communication. It is further divided into five components representing the time spent for EPR pair generation, EPR pair distribution, pre-processing, classical communication, and post-processing.\\
    
    Computation time & The portion of the execution time spent on computation, i.e., gate execution.\\

    Execution time & Total execution time, which is the sum of the communication time and computation time.\\
    
%    Coherence & Based on the experimental results from~\cite{muhonen_nnano14}, the decoherence time of electron spin qubits (without implementing the dynamic decoupling technique for extending qubit lifetime) is $T_2 = 268 \ \mu$s. Assuming an exponential decay of the qubit coherence over time, the qubit coherent state after $t$ $\mu$s is obtained as $e^{-t/T_2}$.\\
    Coherence & Coherence time computed as in~\cite{escofet_qce25}, $C(t)=\mathrm{e}^{-t/T_1} \cdot (\frac{1}{2} \mathrm{e}^{-t/T_2} + \frac{1}{2})$, where $T_1$ (thermal relaxation time) and $T_2$ (dephasing time) are parameters.\\
    \bottomrule
  \end{tabularx}
\end{table}

\section{Experiments}
\label{sec:experiments}
In this section, we demonstrate how \qcomm{} can be used to investigate the impact of various architectural and micro-architectural parameters, as well as circuit properties, on the different components that contribute to execution time. We use both randomly generated circuits and selected benchmarks from QASMBench to cover a wide spectrum of workloads, ranging from unstructured entanglement-intensive circuits to structured applications. This diversity ensures that our evaluation captures different stress points for the communication system.

It is worth noting that the results presented in this section are obtained exclusively through simulation. Since fully cryogenically-controlled modular quantum computers with classical interconnects do not yet exist as physical prototypes, direct hardware validation is not currently possible. However, the goal of \qcomm{} is not to provide cycle-accurate predictions of existing platforms, but rather to serve as a design space exploration tool. In this context, what matters is the relative accuracy of the timing model in capturing performance trade-offs across different architectural and micro-architectural configurations, rather than the absolute precision of execution times. The classical subsystem is modeled with high fidelity, accounting for congestion and MAC protocol effects in the NoC/WiNoC, while the quantum subsystem adopts a parametric additive delay model based on configurable gate and teleportation latencies. This ensures that \qcomm{} can flexibly reflect different technology assumptions as they evolve, while reliably highlighting communication-related bottlenecks.

Table~\ref{tab:parameters} summarizes the parameter values or ranges used throughout the experiments. Link latency and bandwidth values are selected based on parameters reported in the current literature on NoC and WiNoC architectures. These values are complemented with abstractions to enable broad design-space exploration.
\begin{table}
    \centering
    \caption{Physical parameters, architectural and micro-architectural parameters, and circuit properties used for the experiments.}
    \label{tab:parameters}
    \begin{tabular}{lc}
        \toprule
        \multicolumn{2}{c}{Physical parameters and microarchitectural parameters} \\
        \midrule
        Mean of EPR pair generation time    & $10^3$ \si{\nano\second}$^{*}$ \\
        EPR pair distribution time          & 0.01 \si{\nano\second}$^{*}$ \\
        Pre-processing time                 & 390 \si{\nano\second}$^{*}$ \\
        Post-processing time                & 30 \si{\nano\second}$^{*}$ \\
        NoC clock frequency                 & from 10 \si{\mega\hertz} to 1 \si{\giga\hertz} \\
        WiNoC bandwidth                     & 12 Gbps \\
        RAM bandwidth                       & 128 Gbps$^{**}$ \\ 
        Bits per instruction                & 4$^{\dag}$\\
        Decode time                         & $d_1=0$ \si{\nano\second}, $d_2=10$ \si{\nano\second}$^{\ddag}$\\
        \midrule
        \multicolumn{2}{c}{Architectural parameters and circuit properties} \\
        \midrule
        Network topology & mesh \\
        Mesh size & from $1 \times 1$ to $10 \times 10$ \\
        Number of QCs & from 1 to 100 \\
        LTM ports per QC & from 1 to 5 \\
        Physical qubits & from 10 to 10,000 \\
        Link width & from 2 to 10 bits \\
        EPR distribution approach & at the \emph{mid-point}, at \emph{both end-points} \\
        Circuit size -- logical qubits & from 10 to 1000\\
        Circuit size -- gates & from 100 to 10,000 two-input gates\\
        \bottomrule
        \multicolumn{2}{p{0.8\linewidth}}{$^{*}$Data from~\cite{morten_arcmp20}. $^{**}$DDR4 SDRAM. $^{\dag}$The instruction set is assumed to consist of up to 16 instructions. $^{\ddag}$See Eq.~\ref{eqn:decode_time}}
    \end{tabular}
\end{table}

\subsection{Notes on Circuit Generation and Mapping}
In this section, most of the experiments are conducted using randomly generated circuits. A random circuit is created by specifying the number of qubits, the total number of gates, and a probability distribution, where the $i$-th entry defines the fraction of $i$-input gates in the circuit. The circuit is constructed slice by slice: at each step, a gate is randomly selected based on the specified probabilities, and its input qubits are randomly chosen according to the gate’s arity. If the selected input qubits are not already used by another gate in the current slice, the gate is added to the slice and the process continues. Otherwise, the current slice is marked as complete, and a new slice is initiated.

Regarding the mapping of logical qubits to physical qubits, several studies in the literature focus on optimizing different metrics, such as minimizing inter-core communication~\cite{escofet_qce24,escofet_tqc25,russo_date25,russo_qce25}. In most of the experiments presented in this paper, we adopt a baseline (``vanilla'') mapping strategy, where each logical qubit $q_i$ is directly mapped to physical qubit $Q_i$, and $Q_i$ is assigned to quantum core $i \mod M$, where $M$ is the total number of QCs. We also assume all-to-all connectivity within each QC. 
To isolate and analyze the role of inter-core communication, this abstraction avoids conflating intra-core routing overheads with teleportation-related costs, which are the primary focus of this work. In practice, many physical platforms exhibit more constrained connectivity (e.g., linear chains or 2D grids), which would increase the number of SWAPs required for intra-core operations. This would raise the intra-core contribution to execution time and could reduce the \emph{relative} impact of inter-core costs; however, because teleportations remain significantly more expensive than local SWAPs, inter-core communication is still expected to dominate overall performance. Extending \qcomm{} to incorporate realistic intra-core connectivity constraints is currently under development and will be released with the next version of the simulator.
However, in Section~\ref{ssec:real_benchmarks}, we additionally evaluate the impact of optimized mapping by leveraging TeleSABRE~\cite{russo_qce25}, a routing and qubit allocation strategy designed for modular quantum architectures. This allows us to assess the benefits of communication-aware mapping on execution time and teleportation overhead.

In this work, we do not explicitly model qubit reuse enabled by mid-circuit measurement and reset (MCMR) techniques~\cite{decross_phr23,brandhofer_qce23}. While reuse can reduce the overall qubit footprint of a circuit, its impact on execution time strongly depends on technology-specific reset latencies and compiler strategies for qubit remapping. Since our focus is on quantifying communication overhead in modular multi-core systems rather than optimizing qubit resource utilization, we adopt a baseline model where qubits are allocated statically throughout execution. Extending \qcomm{} to incorporate qubit reuse is an interesting direction for future work, particularly to explore trade-offs between reset time, mapping flexibility, and inter-core teleportation costs.

\subsection{Impact of LTM Ports}
First, let us focus on the communication time, which refers to the portion of the execution time spent on inter-core communications. We analyze the impact of the number of LTM ports per QC on communication time in a system consisting of 16 QCs. The classical communication system is modeled as a $4 \times 4$ mesh-based NoC, as illustrated in Figure~\ref{fig:interconnect_options}.

For the quantum communication system, we consider two scenarios:
\begin{enumerate}
    \item \textbf{Single-hop teleportation} -- The EPR generator is directly connected to each QC via point-to-point links, allowing teleportation between any pair of QCs in a single step. This corresponds to the EPR generation at the mid-point approach described in Sec.~\ref{ssec:epr_generator}.
    
    \item \textbf{Multi-hop teleportation} -- EPR pairs are generated only between neighboring QCs in the mesh topology. As a result, teleportation between non-adjacent QCs requires multiple hops through intermediate QCs. This corresponds to the EPR generation at the source or at both end-points approaches described in Sec.~\ref{ssec:epr_generator}.
\end{enumerate}
The NoC operates at a clock speed of 1 GHz and employs 8-bit links. Each QC integrates 10 physical qubits, resulting in a total of 160 physical qubits across the system.

To evaluate communication performance under different traffic conditions~\cite{rached_iscas24}, we utilize three randomly generated circuits, each containing 100 qubits and 1,000 gates. These circuits vary in the ratio of 1-input to 2-input gates. The proportion of 2-input gates significantly impacts communication traffic, as these gates require communication whenever the involved qubits are located in different QCs. The three circuit configurations include: (1) 75\% 1-input gates and 25\% 2-input gates, (2) a balanced 50\%-50\% mix, and (3) 25\% 1-input gates and 75\% 2-input gates.

\begin{figure}
    \centering
    \includegraphics[width=0.8\columnwidth]{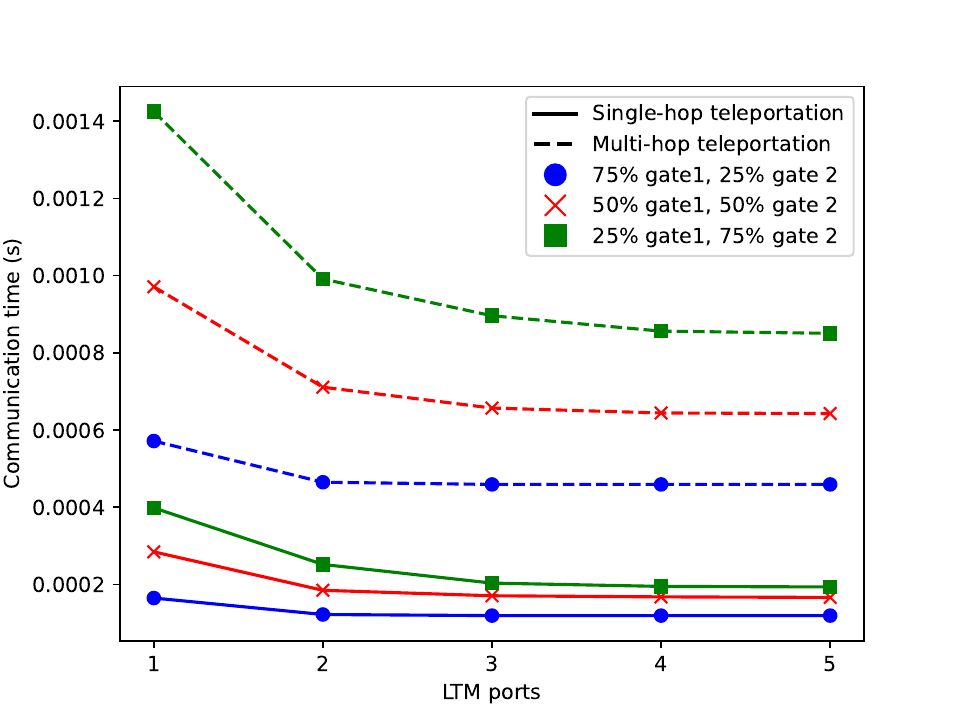}
    \caption{Impact of the number of LTM ports on communication time for three different random circuits (100 qubits, 1,000 gates) characterized by a different ration between 1-input and 2-input gates.}
    \label{fig:commtime_vs_ltmports}
\end{figure}

Figure~\ref{fig:commtime_vs_ltmports} illustrates the total communication time for different LTM port counts. The solid-line curves represent the single-hop teleportation scenario, while the dashed-line curves correspond to multi-hop teleportation, where multiple teleportation steps are required along an XY routing path in the mesh topology when QCs are not directly connected. As expected, circuits with a higher proportion of 2-input gates experience increased communication overhead, as these gates require inter-QC communication, leading to longer communication times. However, as the number of LTM ports increases, communication time decreases due to greater parallelization of data transfer between QCs. Interestingly, while adding LTM ports generally improves communication efficiency, our results show that beyond three LTM ports, no further reduction in communication time is observed.

\begin{figure}
\centering
    \begin{minipage}{0.5\textwidth}
    \centering
    \includegraphics[width=1.0\textwidth]{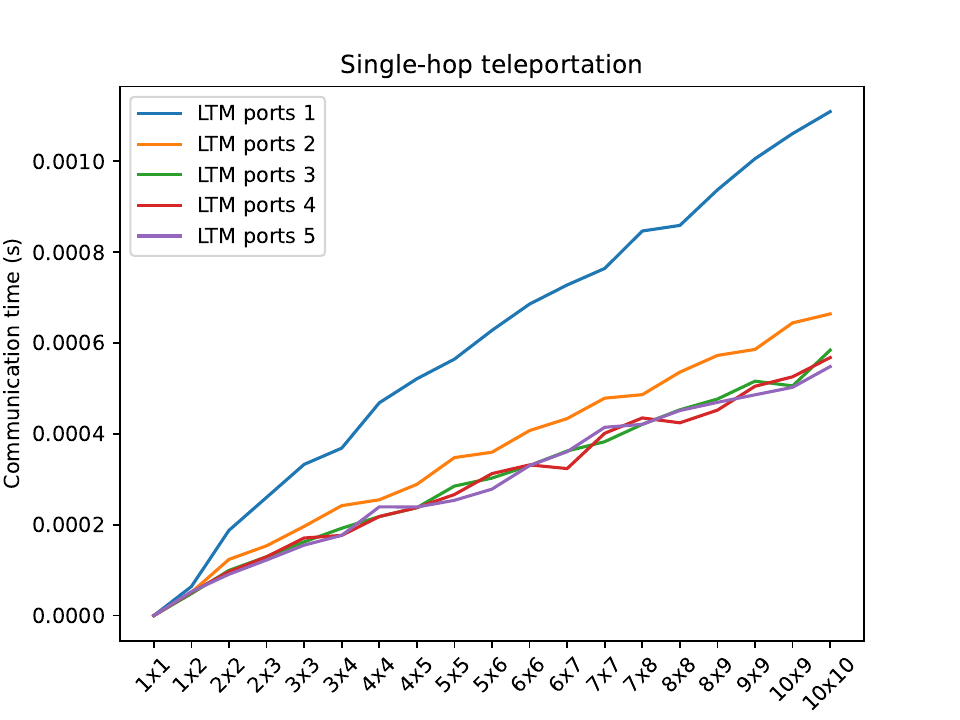}
    \subcaption{}\label{fig:commtime_vs_nocsize-sh}
    \end{minipage}%
    \begin{minipage}{0.5\textwidth}
    \centering
    \includegraphics[width=1.0\textwidth]{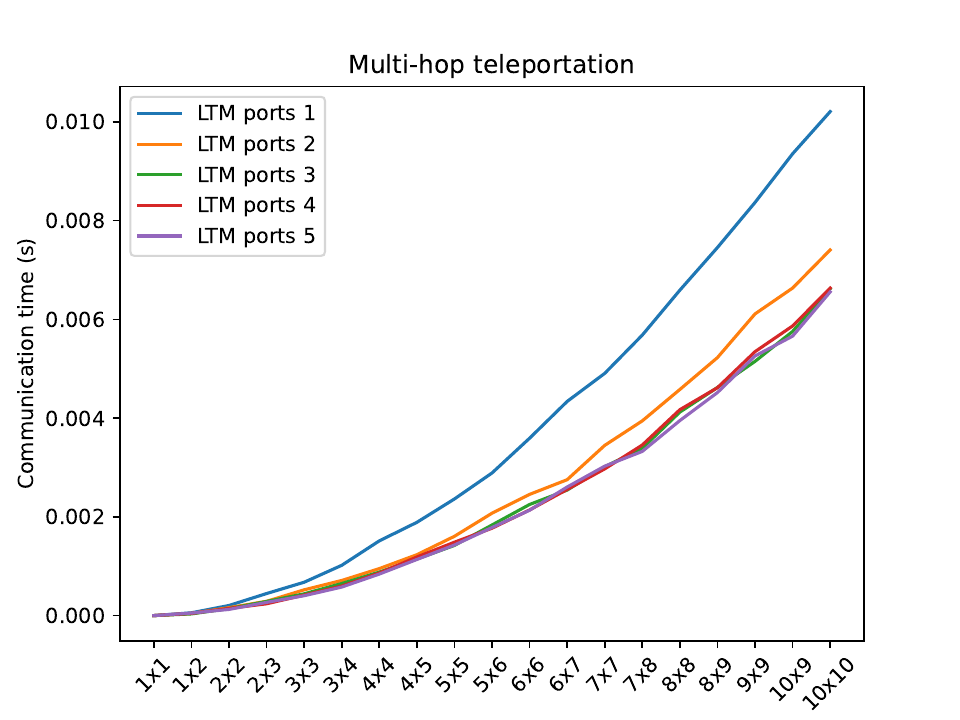}
    \subcaption{}\label{fig:commtime_vs_nocsize-mh}
    \end{minipage}%
\caption{Communication time \emph{vs.} NoC size for different number of LTM ports for single-hop teleportation (\subref{fig:commtime_vs_nocsize-sh}) and multi-hop teleportation (\subref{fig:commtime_vs_nocsize-mh}). For a $n \times m$ mesh, we consider a random circuit with $10n$ logical qubits and $100n$ gates with 50\% 1-input gates and 50\% 2-input gates.}
\label{fig:commtime_vs_nocsize}
\end{figure}

Our findings reveal that communication efficiency saturates after three LTM ports, regardless of network size (i.e., the number of QCs). To validate this, we measured communication time as we scaled up the network size while proportionally increasing circuit complexity. For a baseline single-QC system with 15 physical qubits, the mapped circuit consists of 10 logical qubits and 100 gates. When simulating an $n$-QC system, we map a circuit with $10n$ logical qubits and  $100n$ gates. Figure~\ref{fig:commtime_vs_nocsize} confirms this trend, showing that communication time remains stable beyond three LTM ports, irrespective of the network size.

\subsection{Execution Time Breakdown}
We analyze the impact of various NoC architectural characteristics on overall execution time, aiming to identify the key factors that contribute most significantly to it. Our study is based on a $4 \times 4$ NoC (16 QCs) with 8-bit links, 10-qubit and 2 LTM ports per QC. The mapped circuit is randomly generated and consists of 100 logical qubits and 1,000 two-input gates.

\begin{figure}
    \centering
    \includegraphics[width=0.8\columnwidth]{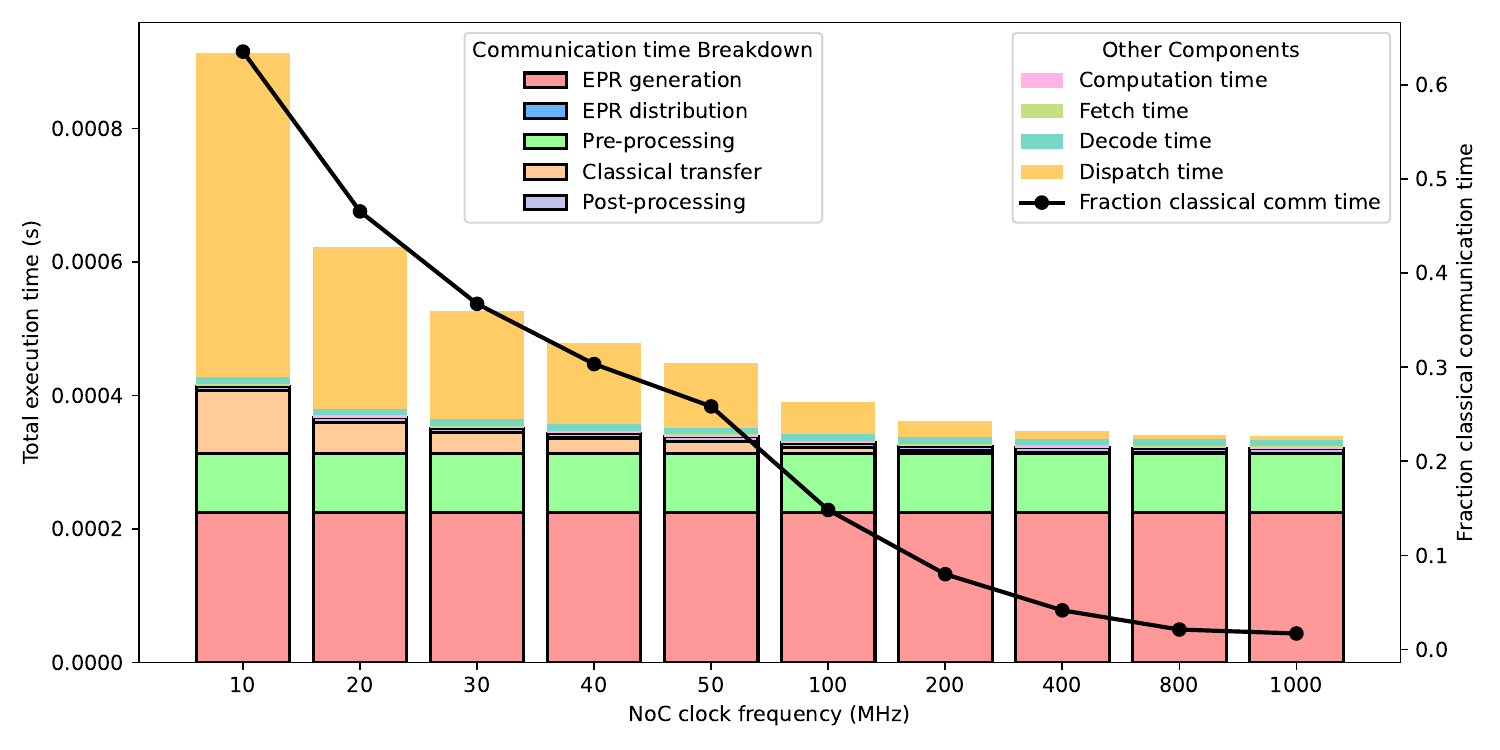}
    \caption{Breakdown of communication time by NoC clock frequency.}
    \label{fig:commtimebd_vs_nocfreq}
\end{figure}

Figure~\ref{fig:commtimebd_vs_nocfreq} presents a breakdown of communication time as the NoC clock frequency varies from 10~MHz to 1~GHz. As observed, communication time--highlighted with a black boundary and encompassing EPR generation and distribution times, pre- and post-processing times, and classical transfer (i.e., the classical communication time introduced by the NoC to support the teleportation protocol)--dominates the overall execution time. Classical communication, which includes the classical transfer component of communication time and the dispatch phase, plays a significant role up to a clock frequency of 50~MHz. However, its impact on total execution time diminishes at 100 MHz and beyond.

\begin{figure}
    \centering
    \includegraphics[width=0.8\columnwidth]{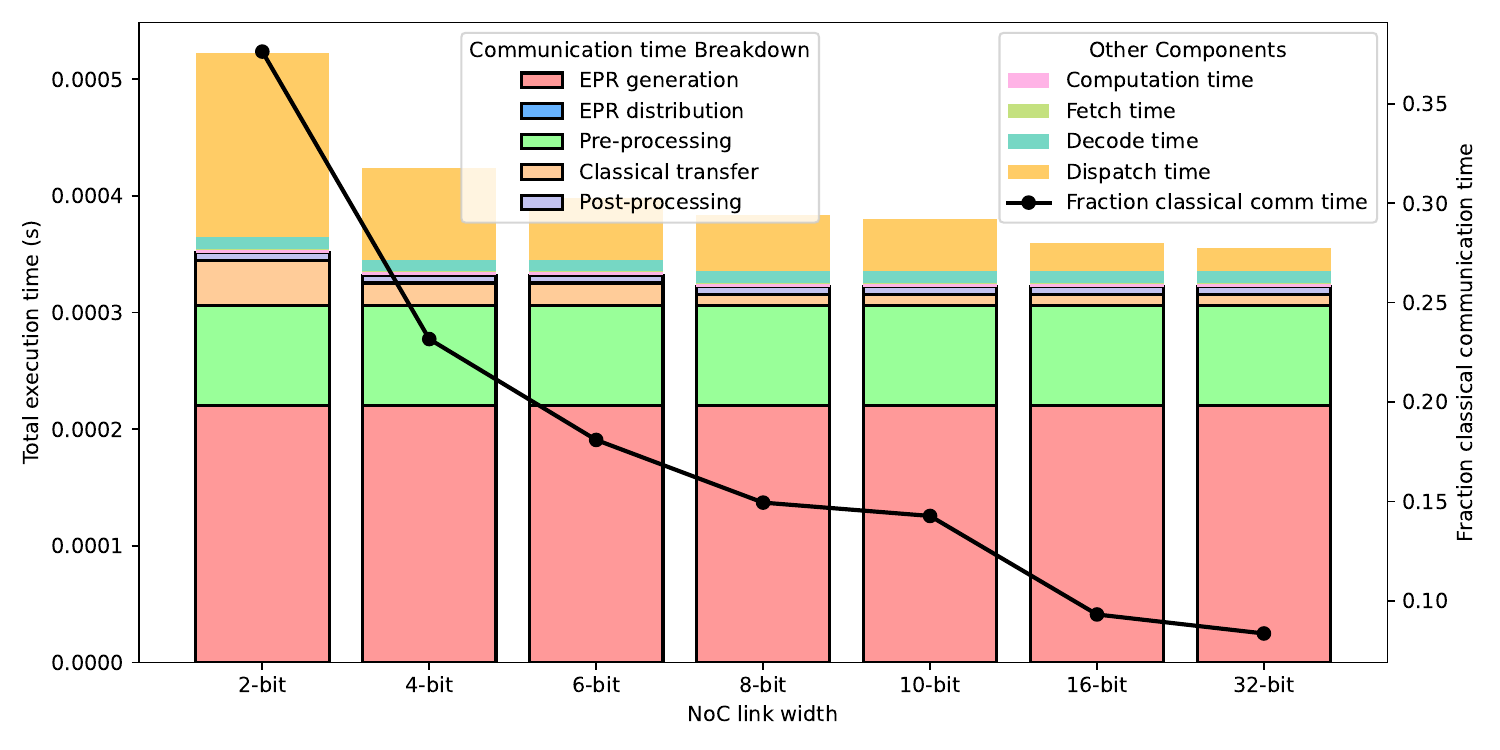}
    \caption{Breakdown of communication time by NoC link width.}
    \label{fig:commtimebd_vs_noclink}
\end{figure}

A similar trend is observed when the NoC link width is varied, as shown in Figure~\ref{fig:commtimebd_vs_noclink}. Although the contribution of classical transfers appears negligible in relation to the communication time, classical communication plays a crucial role in supporting the dispatch phase. Overall, the fraction of classical communication time is less than 15\% for link widths greater than 8 bits.

\begin{figure}
    \centering
    \includegraphics[width=0.8\columnwidth]{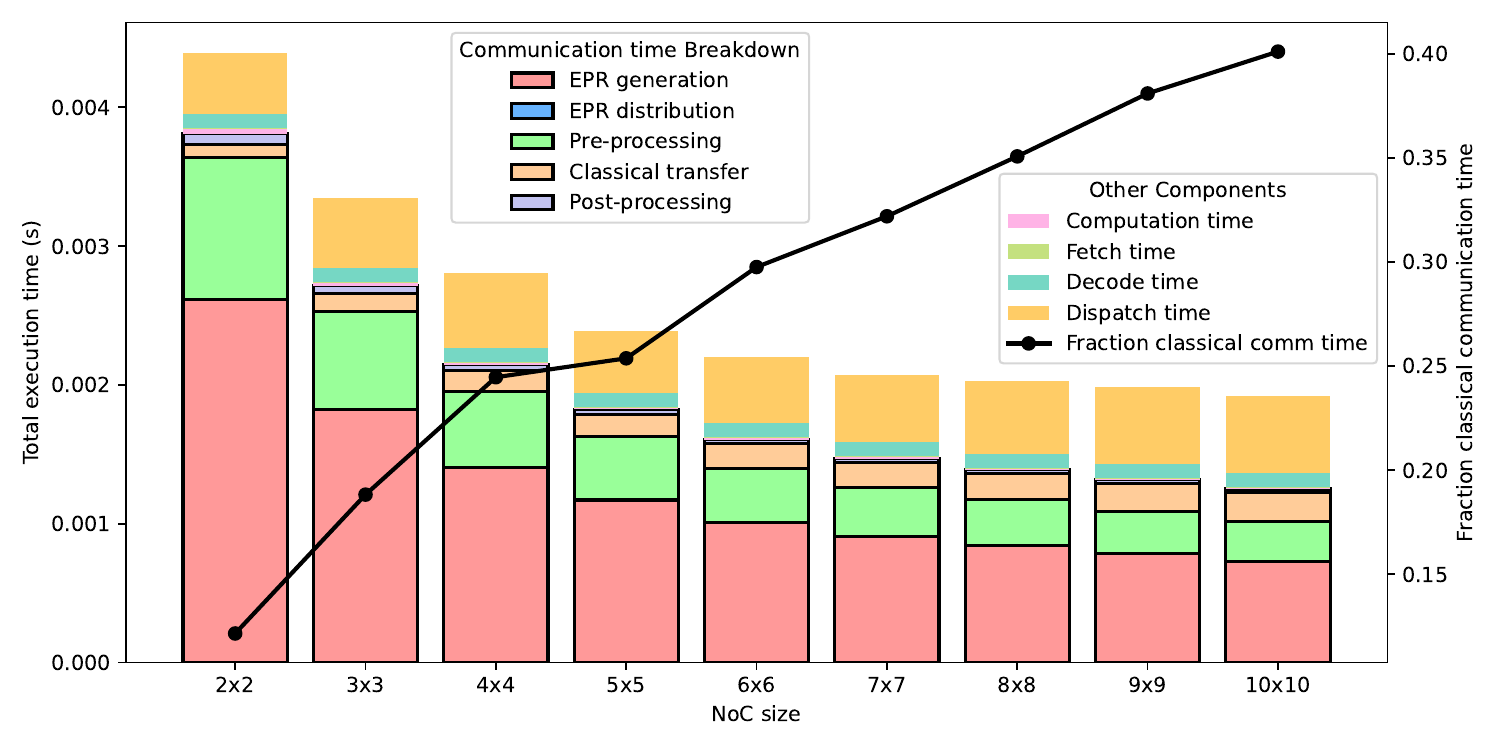}
    \caption{Breakdown of communication time by NoC size.}
    \label{fig:commtimebd_vs_nocsize}
\end{figure}

Finally, Figure~\ref{fig:commtimebd_vs_nocsize} presents a breakdown of communication time across various NoC sizes, using the same random circuit with 1,000 qubits and 10,000 gates. For a system with $n \times n$ QCs, we assigned $\lceil 1000/n^2 \rceil$ physical qubits per core, ensuring that the total number of physical qubits remained constant across all system configurations considered. As expected, total execution time decreases as NoC size increases, benefiting from greater parallelism. However, an interesting trend emerges: the contribution of classical communication becomes more significant with larger NoC sizes. This is because, unlike other factors affecting communication time (e.g., EPR generation and preprocessing), classical communication is directly influenced by the distance between communicating nodes. As the NoC size grows, the average communication distance increases, making classical communication a more prominent cost factor in which its contribution can reach 40\% of the total execution time for the case of a $10 \times 10$ QCs system.

\subsection{Wired vs. Wireless Interconnect}
Classical communication can be implemented using either a wired or wireless communication system, namely NoC or WiNoC. To evaluate the impact of one choice over the other on execution time, we consider a $10 \times 10$ system, where each QC contains 20 physical qubits, resulting in a total of 2,000 physical qubits. Each core has a single LTM port, and we map a circuit consisting of 1,000 qubits and 10,000 gates (60\% 1-input and 40\% 2-input gates).

\begin{figure}
    \centering
    \includegraphics[width=0.8\columnwidth]{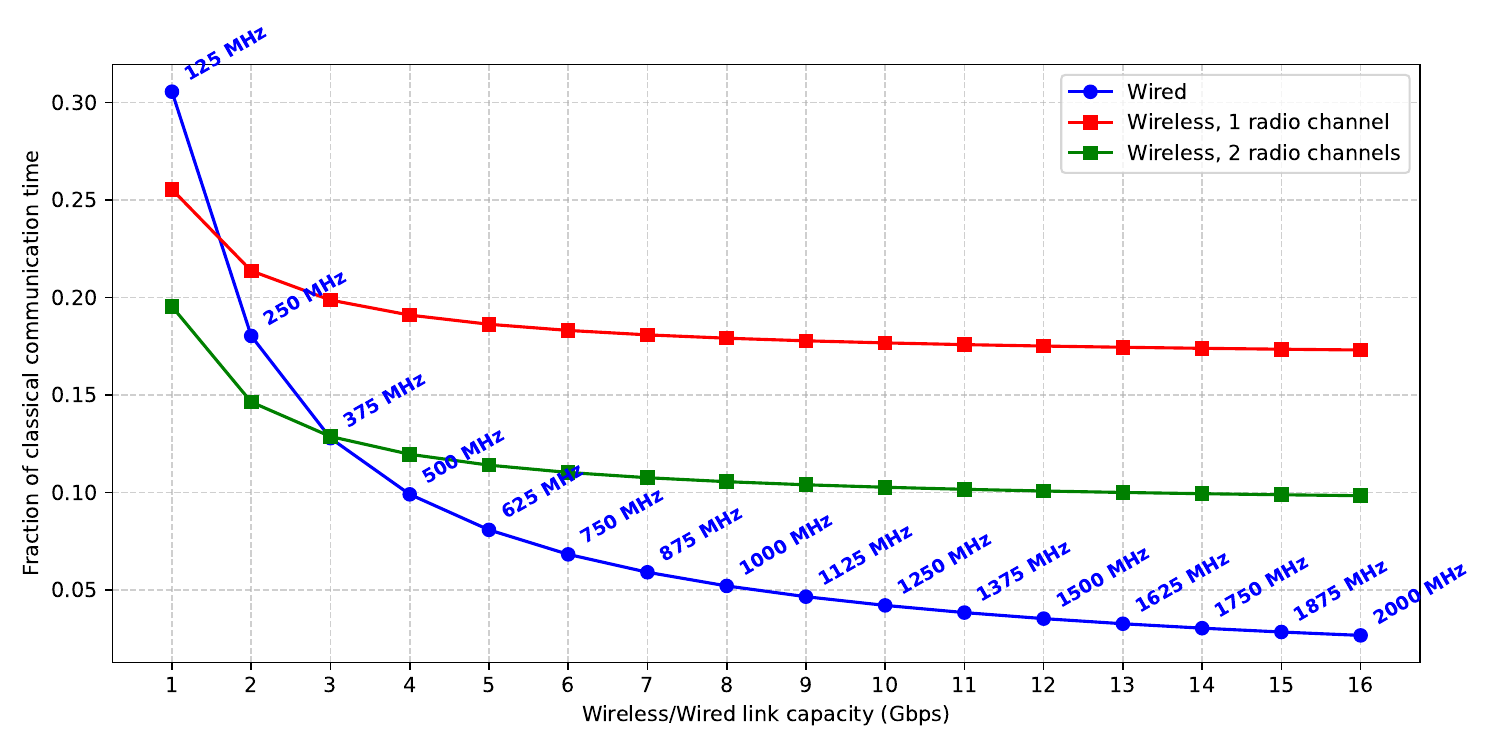}
    \caption{Impact of wired (NoC) and wireless (WiNoC) communication on execution time. }
    \label{fig:wired_vs_wireless}
\end{figure}

We gradually increase the wireless/wired link capacity from 1 to 16 Gbps and measure the fraction of execution time attributed to classical communication, as shown in Figure~\ref{fig:wired_vs_wireless}. In the NoC configuration, links are 8 bits wide, and link capacity is adjusted by varying the router clock frequency. For the WiNoC setup, we consider two cases: one with a single radio channel and another with two radio channels.

The first key observation from the graph is that NoC outperforms WiNoC in communication performance for clock frequencies above 375~MHz. Additionally, when using WiNoC, the contribution of classical communication to execution time does not decrease beyond 17\% and 10\% for the one-channel and two-channel cases, respectively. This limitation arises from the MAC protocol used, which relies on token circulation among the WIs~\cite{ditomaso_tpds15,deb_tc13,palesi_jlpa15}. A WI can only transmit on the radio channel when it holds the token, restricting overall efficiency. Developing more advanced MAC protocols would be an important research direction.

It is important to note that increasing the NoC clock frequency comes at a cost, as it directly impacts power dissipation, which is a critical factor in this context. The current version of the simulator does not yet include a power consumption model, and analyzing the trade-off between performance and power consumption is left for future development.

\begin{figure}
    \centering
    \includegraphics[width=0.8\columnwidth]{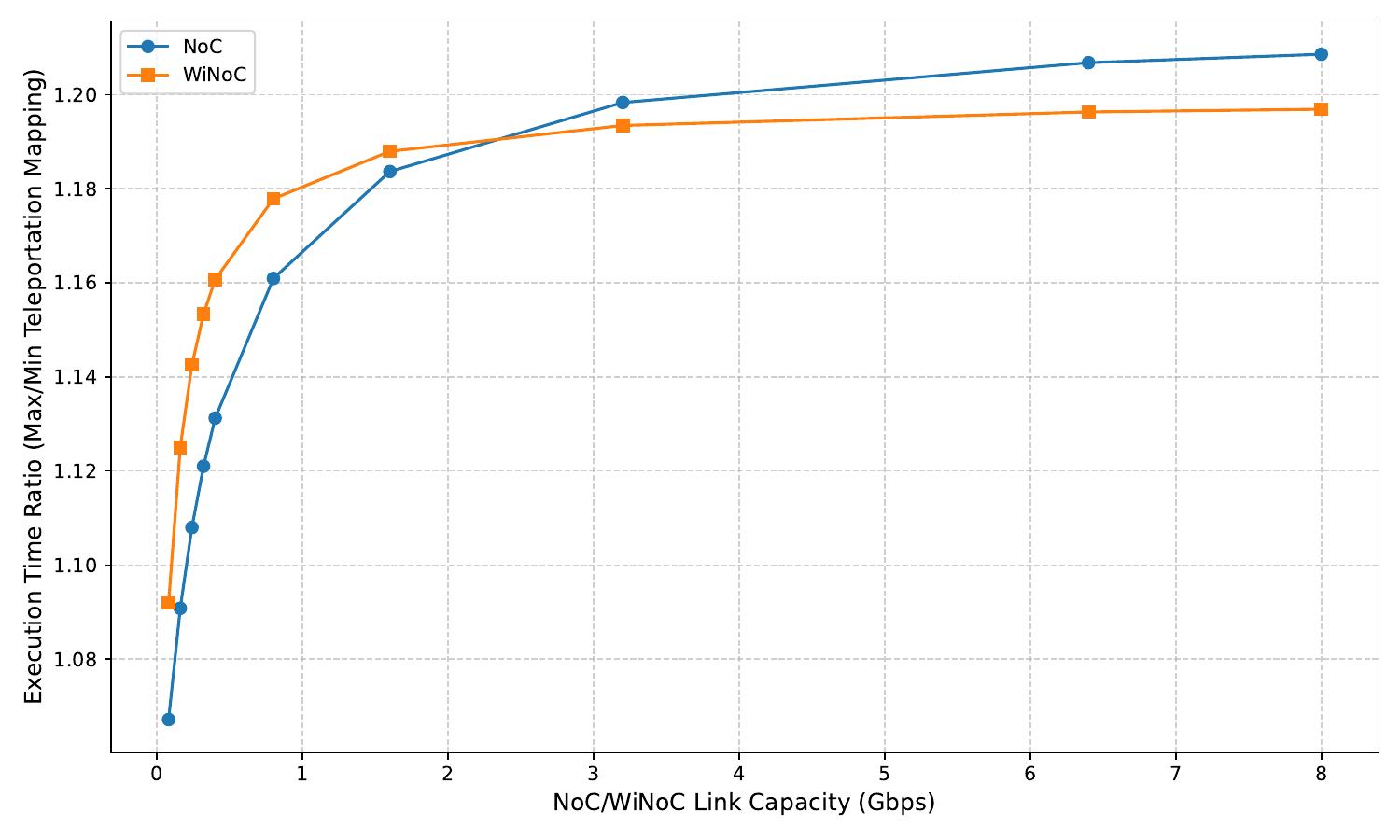}
    \caption{Execution time ratio between mappings with maximum and minimum teleportation counts under varying link capacity for NoC and WiNoC interconnects.}
    \label{fig:min_max_teleportations}
\end{figure}

To further quantify the sensitivity of execution time to classical communication, we evaluated a 64-QC system (20 qubits per QC) executing a 1024-qubit circuit with 1,024 gates (40\% single-qubit, 60\% two-qubit). From 1,000 random mappings, we selected those yielding the minimum and maximum teleportation counts and simulated both under NoC and WiNoC interconnects while varying link capacity. The results in Figure~\ref{fig:min_max_teleportations} show that the execution time ratio between maximum- and minimum-teleportation mappings increases with link capacity, confirming that classical communication becomes increasingly critical as system performance is optimized; notably, WiNoC exhibits higher sensitivity at low bandwidths, whereas NoC dominates at higher bandwidths.

\subsection{Real Benchmarks Analysis}
\label{ssec:real_benchmarks}
In this subsection, we evaluate a diverse set of benchmark circuits drawn from the Qiskit framework~\cite{qiskit2024} and MQTBench~\cite{quetschlich2023mqtbench}. The benchmarks include: amplitude estimation (ae), Greenberger-Horne-Zeilinger state preparation (ghz), graph state preparation (graphstate), quantum Fourier transform (qft), quantum neural network (qnn), and a randomly generated circuit. All circuits were optimized, transpiled, and decomposed using Qiskit to target a native gate set composed of Z-axis rotations, square root of NOT, bit-flip, and controlled bit-flip gates.

Each circuit is generated with 25 qubits and mapped onto a multi-core quantum system composed of 4 QCs, each with 9 physical qubits. The mapping process is performed using TeleSABRE~\cite{russo_qce25}, a quantum routing framework for multi-core quantum processors that extends the SABRE~\cite{li_asplos19} heuristic. TeleSABRE aims to reduce both inter-core communication overhead and the number of intra-core SWAP operations, thereby enabling more efficient circuit execution through improved support for teleportation and local gate execution.

\begin{figure}
    \centering
    \includegraphics[width=0.8\columnwidth]{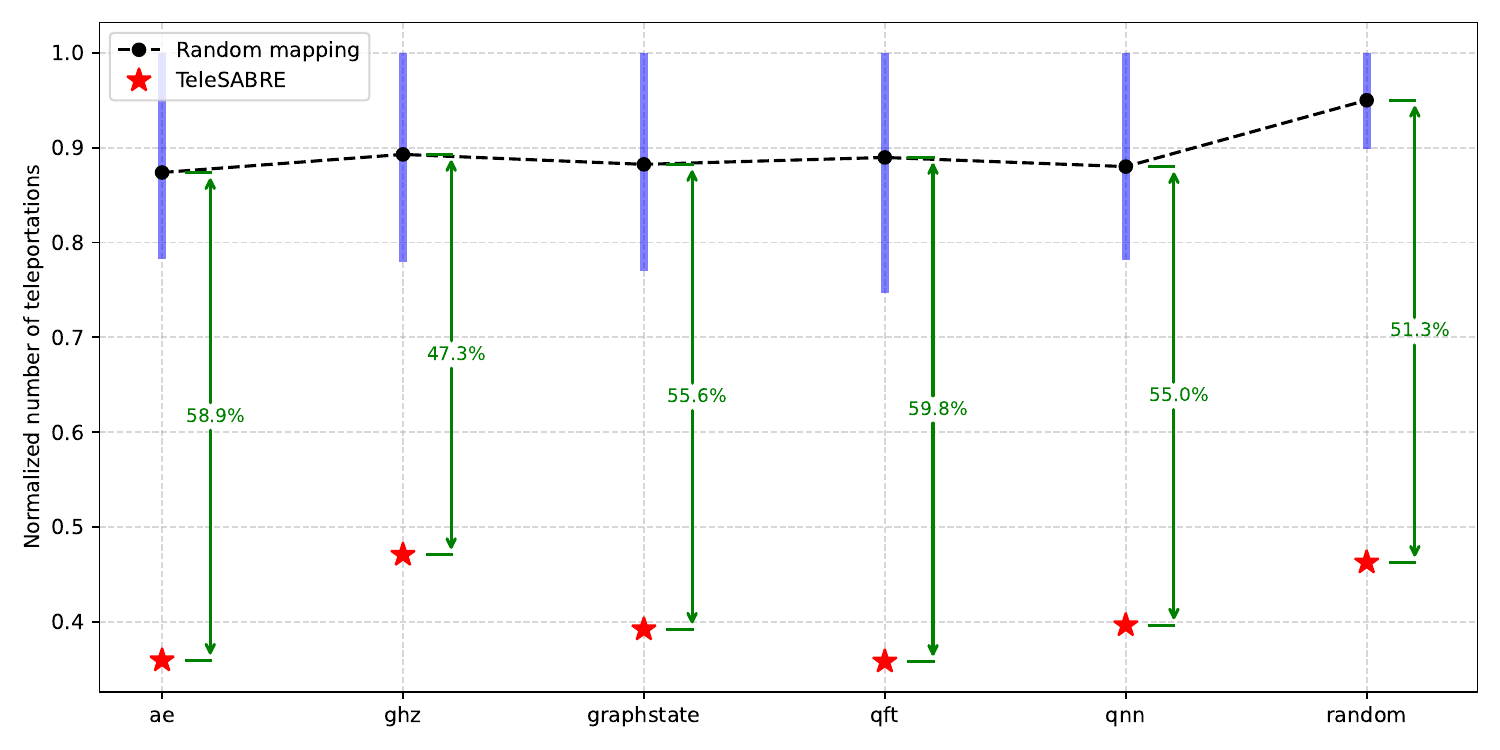}
    \caption{Normalized number of teleportations for randomly generated mappings and the optimized mapping produced by TeleSABRE~\cite{russo_qce25}. For the random case, 100 mappings are evaluated, and the minimum, maximum, and average teleportation counts are shown.}
    \label{fig:tp_rnd_vs_telesabre}
\end{figure}

The impact of the mapping strategy is illustrated in Figure~\ref{fig:tp_rnd_vs_telesabre}, which reports the number of teleportations required when using TeleSABRE compared to a baseline with random logical-to-physical qubit assignments. Specifically, we consider 100 random mappings, and for each, we record the normalized minimum, maximum, and average number of teleportations---shown as a purple bar and a black dot, respectively. These results use the \emph{load-aware} destination selection mode (see Table~\ref{tab:arch_params}). The red star indicates the normalized number of teleportations observed when using TeleSABRE, which employs the \emph{load-independent} destination selection mode, as the mapper itself specifies the destination QC for each teleportation. As shown in the figure, using a dedicated mapping strategy like TeleSABRE can, on average, halve the number of teleportations, leading to significant improvements in execution performance.

\begin{figure}
    \centering
    \includegraphics[width=0.9\columnwidth]{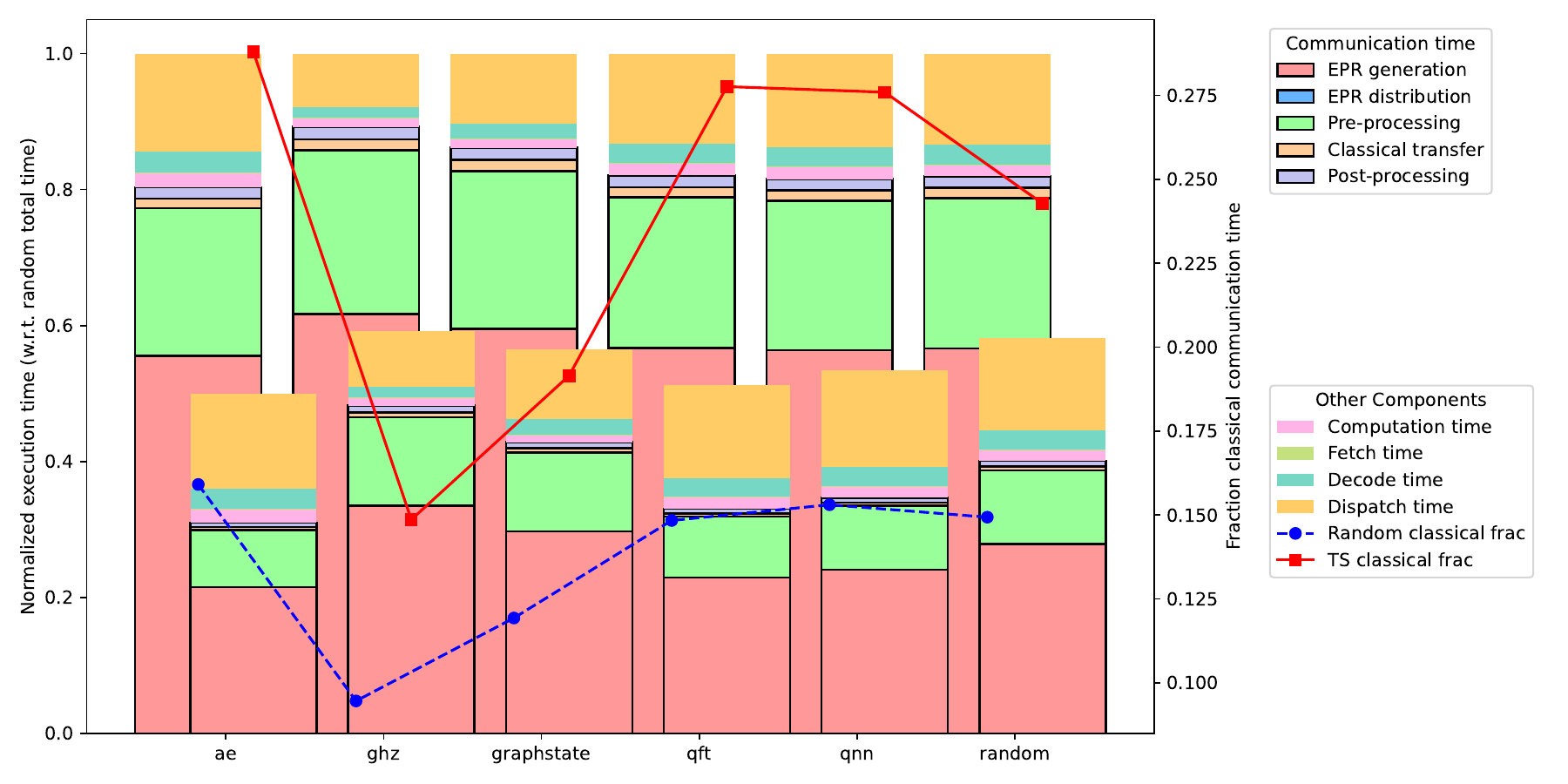}
    \caption{Breakdown of normalized execution time for circuits mapped using random and the optimized mapping produced by TeleSABRE~\cite{russo_qce25}. Execution time is normalized with respect to the random mapping.}
    \label{fig:bd_rnd_vs_telesabre}
\end{figure}

Figure~\ref{fig:bd_rnd_vs_telesabre} shows the breakdown of execution time for both a random mapping and the optimized mapping produced by TeleSABRE. Execution times are normalized with respect to the random mapping. As shown, the reduction in teleportations achieved by TeleSABRE leads to a corresponding decrease in total execution time---averaging over 50\% compared to the random mapping. An interesting observation concerns the fraction of execution time spent on classical communication, which becomes more significant in the optimized case. As execution time decreases due to more efficient mapping, the relative contribution of classical communication increases---from approximately 15\% to 30\%---compared to the random mapping case, where it remains within a 10\% to 15\% range.

\subsection{Projected Trends}
Quantum computing technology is advancing at a rapid pace. For instance, the coherence times of superconducting quantum computers have increased from just 1 nanosecond to 100 microseconds over the past decade~\cite{devoret_science13}. Moreover, current superconducting qubits continue to show an improving trend in coherence times~\cite{devoret_science13}.

The previous analysis of the communication system's role is based on present-day quantum technology parameters, as summarized in Table~\ref{tab:parameters}. Under these conditions, classical communication generally plays a secondary role in the total communication time. However, exceptions arise when certain parameters are pushed to their limits---for example, when the NoC link width or clock frequency is extremely low, when the NoC size is very large, or in real benchmark scenarios where highly parallel gate execution causes dispatch and decode operations to dominate.

\begin{figure}
    \centering
    \includegraphics[width=0.8\textwidth]{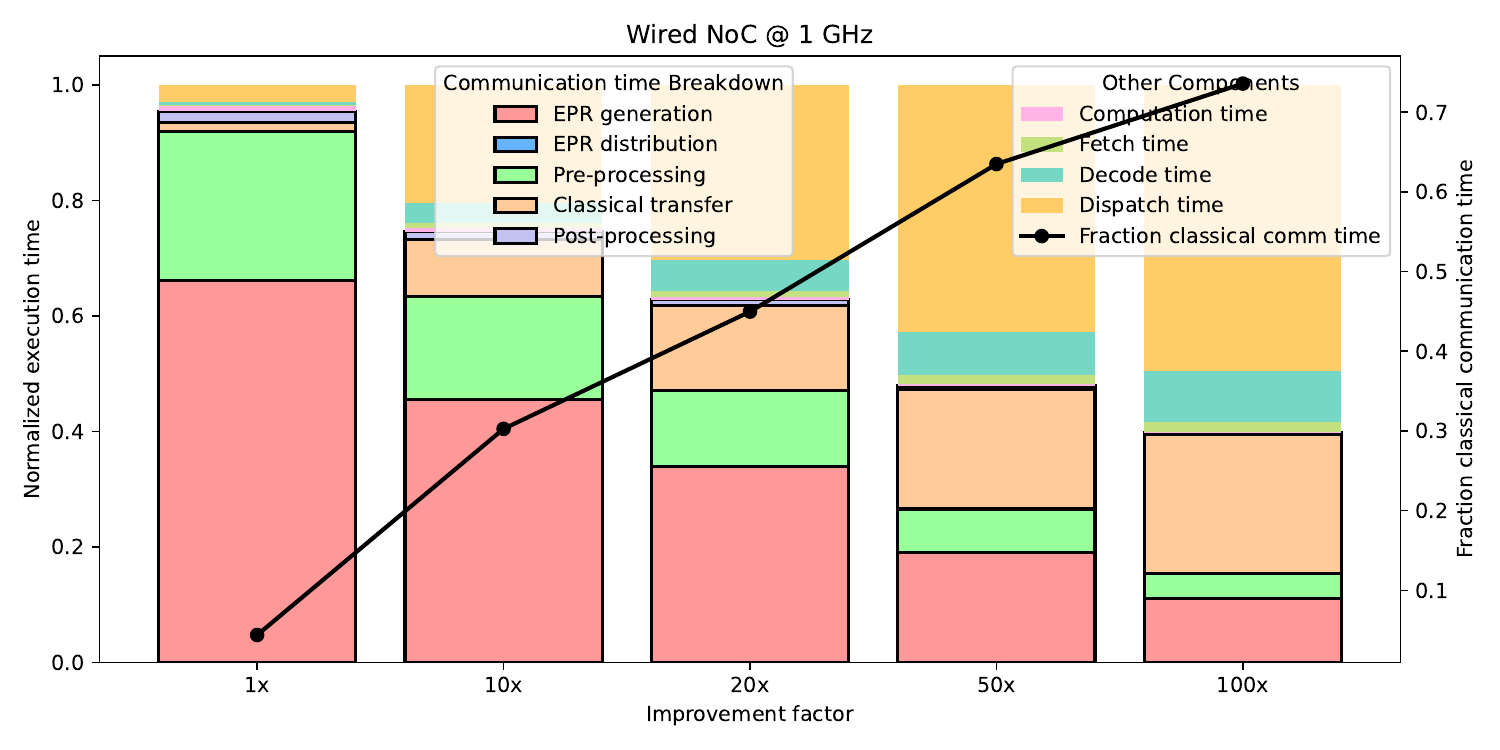}
    \\
    \includegraphics[width=0.8\textwidth]{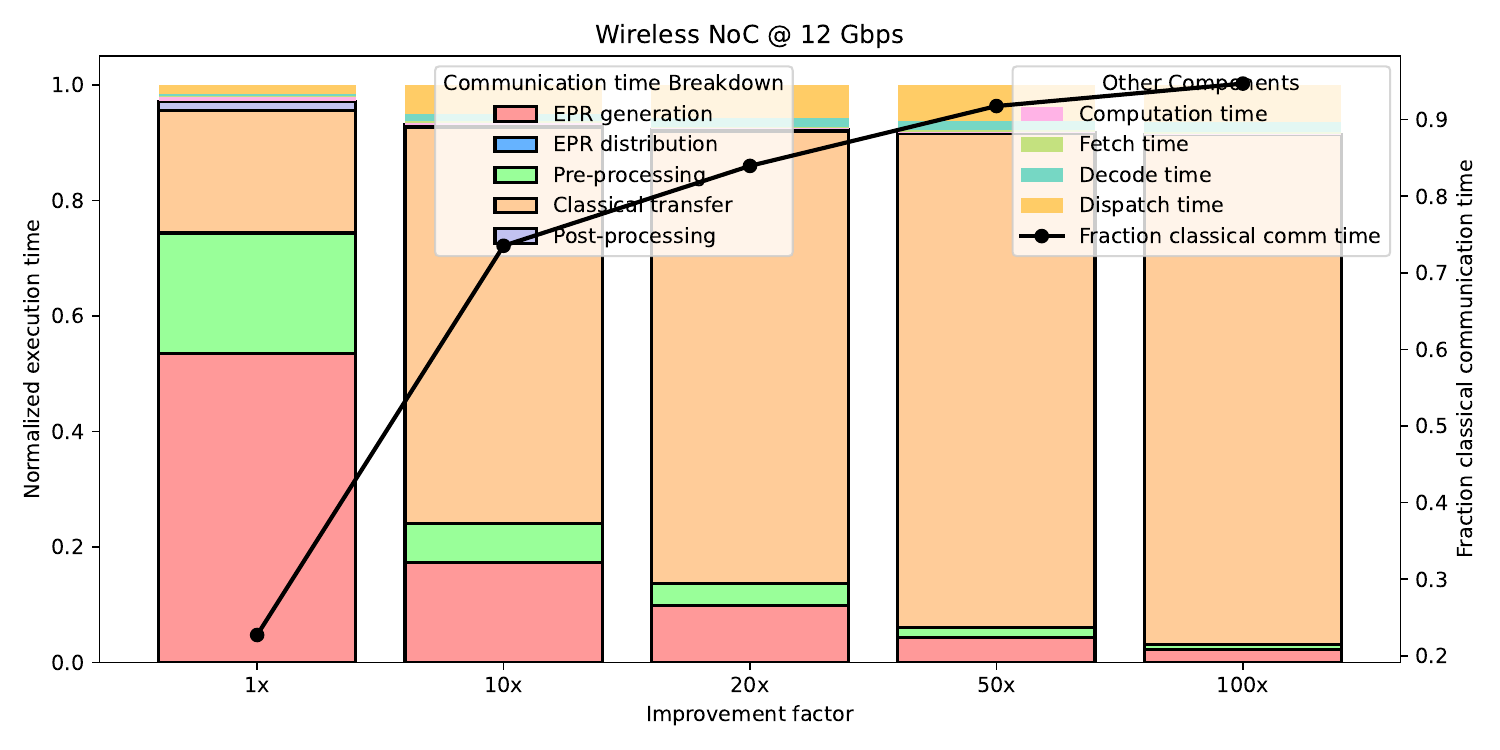}
    \caption{Execution time breakdown as quantum parameters scale by a given improvement factor.}
    \label{fig:previsional_trends}
\end{figure}

To better understand when classical communication becomes a bottleneck, we systematically scale key quantum-related parameters---EPR generation time, pre-processing time, and post-processing time---by the same factor. Our goal is to determine at what point classical communication becomes the dominant limiting factor. We evaluate this by considering two system configurations: one based on a NoC and the other on a WiNoC. Both architectures feature 100 QCs, each with 15 qubits, running a randomly generated circuit consisting of 1,000 qubits and 10,000 CNOT gates. The NoC configuration utilizes 8-bit links and operates at a 1~GHz clock frequency, while the WiNoC configuration employs a single radio channel at 12~Gbps. Figure~\ref{fig:previsional_trends} presents a breakdown of execution time and the fraction attributed to classical communication.

For the NoC-based system, inter-core communication time decreases as quantum technology improves. However, classical communication becomes increasingly significant, surpassing 40\% of the execution time at a 20x improvement factor. The impact of classical communication is even more pronounced in the WiNoC scenario. Unlike the NoC case, inter-core communication time remains largely unchanged with technological improvements, as it is dominated by classical data transfers. The high number of QCs and frequent inter-core communications in randomly generated circuits put significant strain on the single shared radio channel. This issue is further exacerbated by the unicast nature of communication, which is not optimal for wireless transmission, and by the token-based MAC protocol. In this protocol, each QC must wait for the token to circulate among all 100 QCs before it is granted permission to transmit. As shown in the results, even a 10x improvement in quantum technology parameters causes classical communication to become a bottleneck, contributing to over 70\% of the total execution time.

Overall, while classical communication currently plays a minor role compared to other components, this may change as quantum technology continues to advance. Future improvements could make classical communication the primary bottleneck, underscoring the need for further research in this area.

\subsection{Runtime Analysis}
This section reports the runtime of the simulator as both circuit size and system size scale. All experiments are executed on a laptop equipped with an Apple M2 chip running at 3.5~GHz.

\begin{figure}
\centering
    \begin{minipage}{0.5\textwidth}
    \centering
    \includegraphics[width=1.0\textwidth]{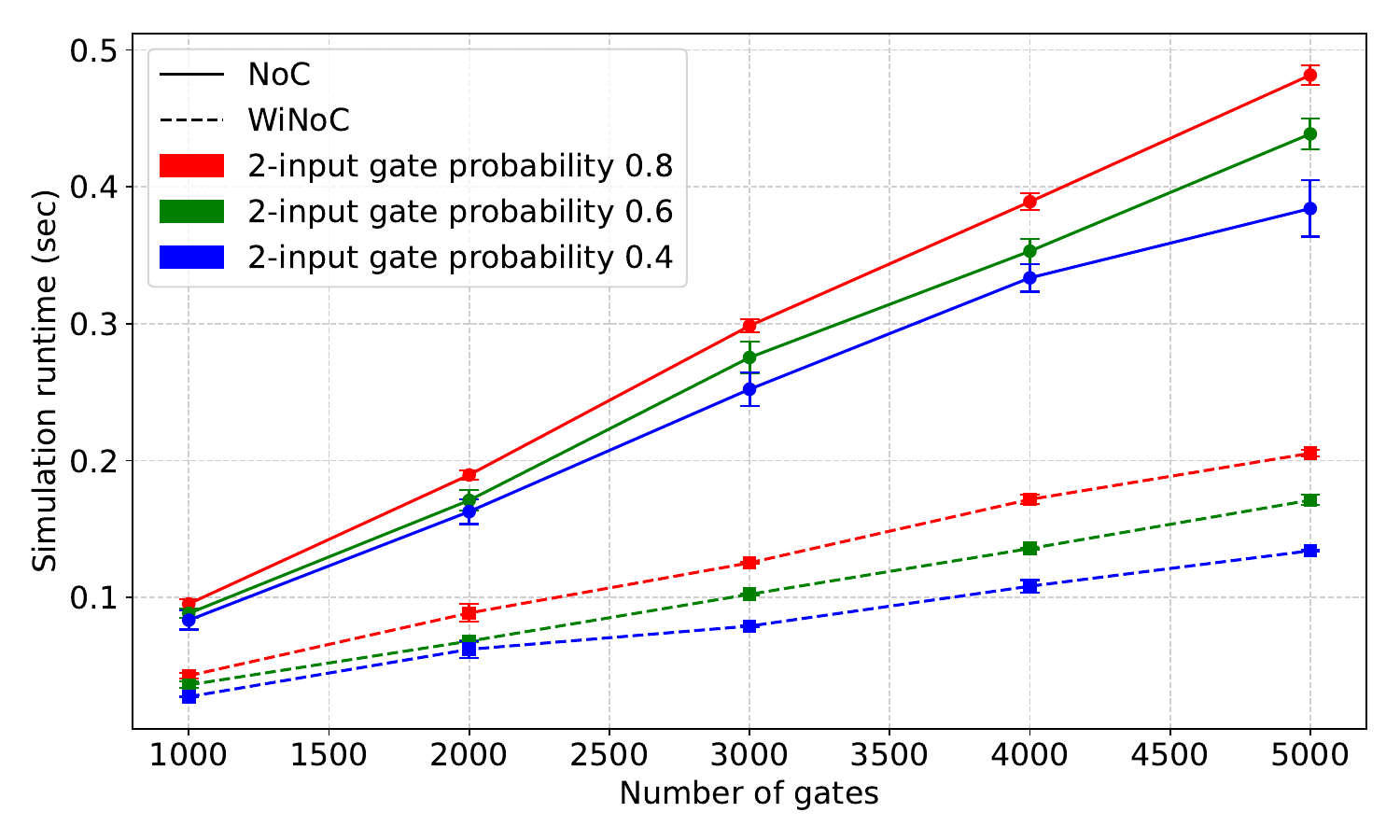}
    \subcaption{}\label{fig:runtime_cs}
    \end{minipage}%
    \begin{minipage}{0.5\textwidth}
    \centering
    \includegraphics[width=1.0\textwidth]{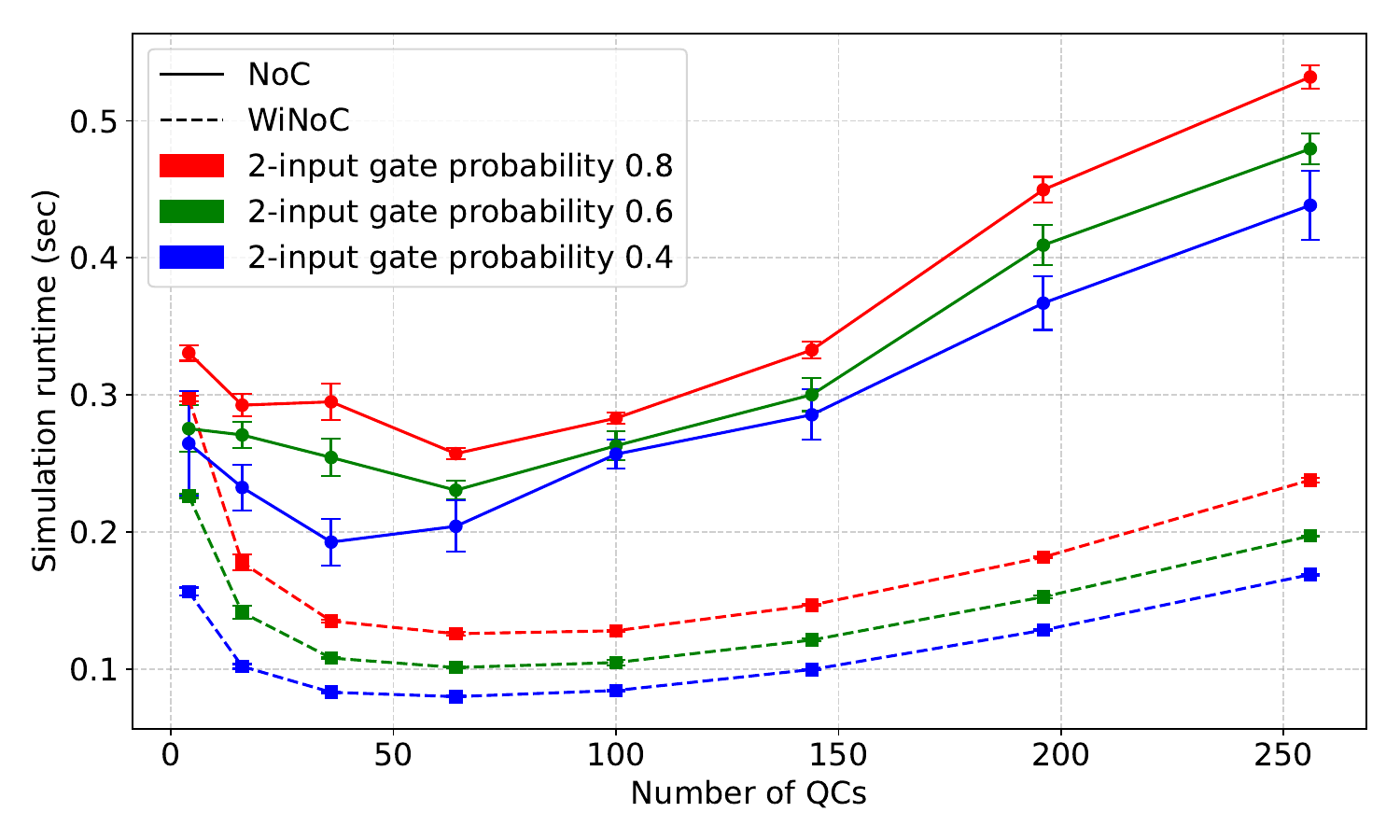}
    \subcaption{}\label{fig:runtime_ss}
    \end{minipage}%
\caption{Runtime of the simulator under different scaling conditions. (\subref{fig:runtime_cs}) Effect of the number of gates on runtime. (\subref{fig:runtime_ss}) Runtime scalability with increasing system size. Error bars indicate 95\% confidence intervals.}
\label{fig:runtime}
\end{figure}
Figure~\ref{fig:runtime_cs} shows the runtime for random input circuits with 1,000 qubits as the number of gates varies. Two system configurations are considered: one using a NoC-based classical interconnect and one using a WiNoC-based interconnect. In both cases, the system comprises 100 QCs with 16 qubits per core. For each gate count, three different circuits are generated by varying the fraction of two-input versus one-input gates. As observed, even for large circuits and system sizes, the simulation runtime remains below 0.5~s. The NoC configuration is consistently more time-consuming than the WiNoC configuration, primarily due to the additional cost of computing routing paths for inter-core communication. Moreover, runtime increases with the fraction of two-input gates, which naturally induce more teleportations and hence more inter-core communication events to be simulated.

Figure~\ref{fig:runtime_ss} analyzes runtime scalability with respect to system size, i.e., the number of QCs. The total number of physical qubits is fixed to 2,000, and thus the number of qubits per core decreases as the number of QCs increases. The results exhibit a non-monotonic trend: runtime decreases initially with increasing core count, then increases beyond a certain point. This behavior reflects a trade-off between parallelism and communication complexity. Increasing the number of cores enhances parallelism, enabling multiple teleportations to be executed concurrently and thus reducing the number of simulation cycles. However, larger core counts also introduce more complex inter-core communication patterns, which eventually dominate and increase runtime.

Overall, these results highlight that \qcomm{} simulations remain lightweight even at large scales (up to thousands of qubits and hundreds of cores), with runtimes on the order of fractions of a second on commodity hardware. This efficiency makes the tool practical for extensive design-space exploration studies.

\section{Conclusion}
\label{sec:conclusion}
In this paper, we introduced \qcomm, an open-source simulator for evaluating the role of classical communication in modular multi-core quantum architectures. Unlike low-level circuit simulators or network-protocol frameworks, \qcomm{} provides an architectural abstraction that integrates quantum teleportation, instruction dispatch, and classical interconnect modeling in cryogenically controlled environments. Through a series of experiments with synthetic and real quantum benchmarks, we showed that while classical communication is not the dominant contributor to execution time in current technology scenarios, it becomes increasingly relevant as systems scale, quantum hardware improves, or optimized qubit-to-core mappings are applied. These findings highlight the need to co-design quantum algorithms, mapping strategies, and interconnect architectures in order to achieve scalable quantum computing.

Despite these contributions, our work has several limitations that also open opportunities for future research. First, the current version of \qcomm{} does not model quantum error correction (QEC), which is expected to introduce significant overhead in both communication and control. Extending the simulator to incorporate error-corrected logical qubits, including schemes such as the surface code, would enable more realistic evaluations.

Second, while our experiments demonstrate scalability up to large multi-core systems, the simulator itself could be further optimized to support very-large-scale (>1000 cores) studies, possibly leveraging parallel or distributed simulation techniques.

Third, \qcomm{} currently focuses on teleportation-based communication; exploring alternative paradigms such as direct state transfer or remote gate execution, and their interplay with classical interconnect latency, remains an open direction.

Fourth, while our evaluation primarily focuses on latency, it is important to note that power consumption is a critical factor in cryogenically integrated quantum systems. Classical interconnects (NoC/WiNoC), control units, and repeated entanglement generation all contribute to the thermal load that must be managed at cryogenic temperatures, where cooling power is extremely limited. In particular, wireless interconnects may reduce wiring overhead but require additional transceiver circuitry, whereas wired NoCs can increase routing congestion and heat dissipation within the cryostat. Modeling these trade-offs is beyond the current scope of \qcomm{}, but we view energy and thermal analysis as a natural extension.

Fifth, \qcomm{} cannot yet be directly validated against hardware, since fully integrated modular quantum computers are still in early research stages~\cite{gent_spectrum25}. Nonetheless, by relying on well-established classical interconnect models and parametric quantum delay assumptions, the simulator offers a valuable tool for relative performance exploration. An important direction for future work is to calibrate the simulator with empirical data as modular prototypes emerge, thereby refining its fidelity while preserving its utility for early-stage design space exploration.

Finally, integration with existing quantum software stacks (e.g., Qiskit, t$|$ket$\rangle$, Cirq) and with network-level simulators (e.g., NetSquid, SeQUeNCe) represents a promising avenue to provide full-stack, communication-aware design exploration. By addressing these limitations, future versions of \qcomm{} will support broader studies on the co-optimization of compilers, interconnects, and error correction, ultimately advancing the design of scalable modular quantum systems.

\bibliographystyle{ACM-Reference-Format}
\bibliography{bibliography}

\appendix

\section{EPR Generation and Distribution}
\label{pdx:epr}
\subsection{Impact on the Architecture}
\begin{figure}
\centering
    \begin{minipage}{0.8\textwidth}
    \centering
    \includegraphics[width=1.0\textwidth]{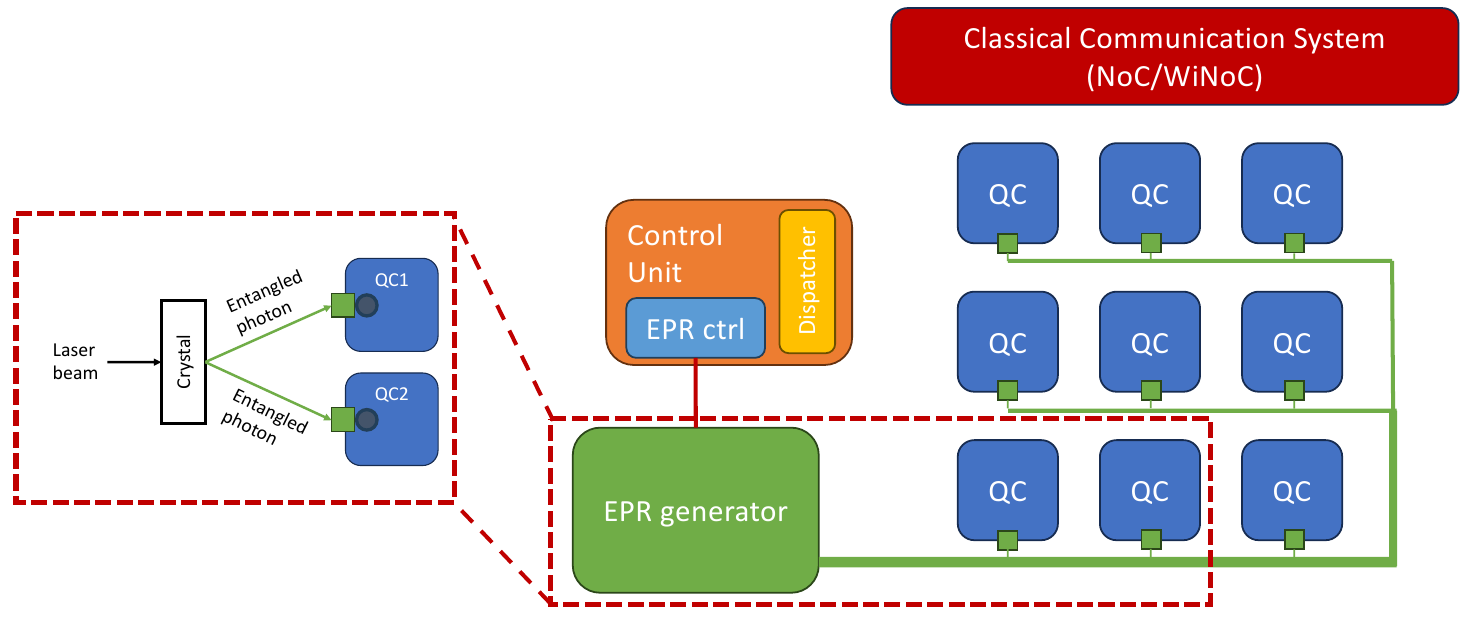}
    \subcaption{System organization when EPR generation and distribution at the mid-point is used.}\label{fig:arch_midpoint}
    \end{minipage}%
    \\
    \begin{minipage}{0.8\textwidth}
    \centering
    \includegraphics[width=1.0\textwidth]{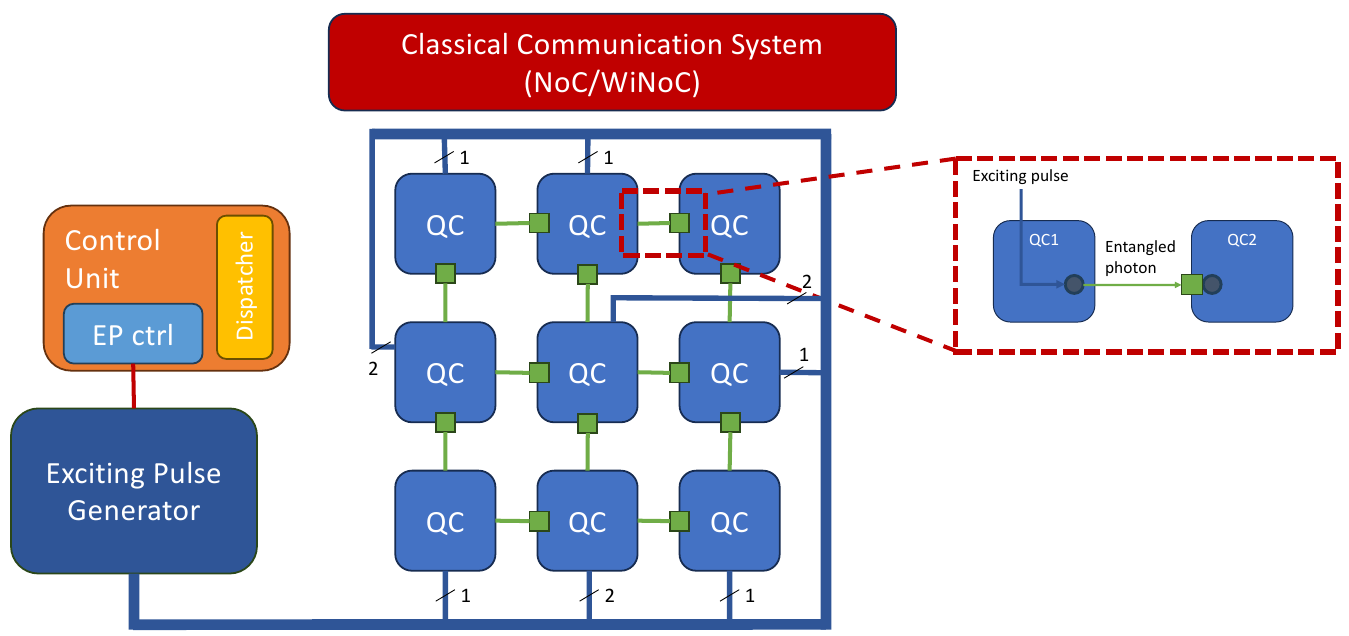}
    \subcaption{System organization when EPR generation and distribution at source is used.}\label{fig:arch_source}
    \end{minipage}
    \\
    \begin{minipage}{0.8\textwidth}
    \centering
    \includegraphics[width=1.0\textwidth]{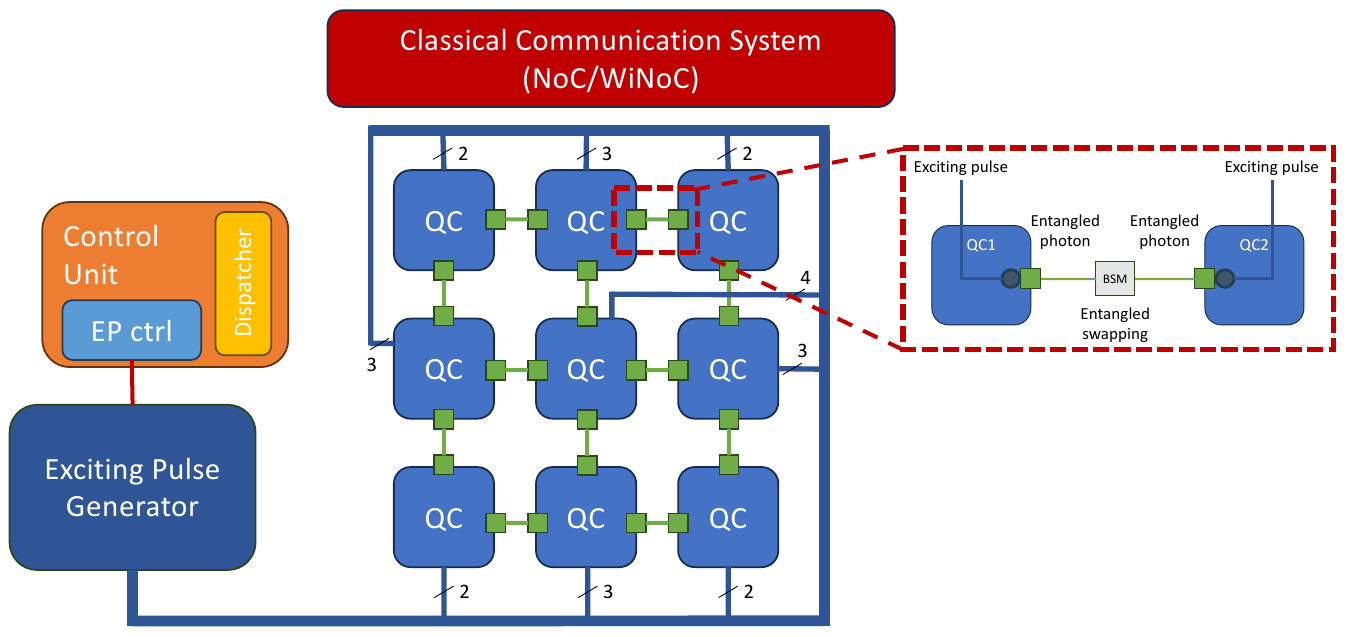}
    \subcaption{System organization when EPR generation and distribution at both end-points is used.}\label{fig:arch_endpoints}
    \end{minipage}
\caption{The three different approaches to EPR generation and distribution result in distinct system architectures.}\label{fig:arch_variations}
\end{figure}

The three different approaches to EPR generation and distribution result in distinct system architectures. This is illustrated in Figure~\ref{fig:arch_variations}, which shows how the system architecture changes depending on the chosen EPR generation and distribution method. So far, we have considered EPR generation and distribution at the midpoint (Figure~\ref{fig:arch_midpoint}). When adopting the other approaches, the EPR Generator module is replaced by the Exciting Pulse Generator module, as shown in Figs.~\ref{fig:arch_source} and~\ref{fig:arch_endpoints}.

Another key difference among these system architectures lies in the interconnection between the EPR Generator/Exciting Pulse Generator and the QCs. In the first approach, a point-to-point connection delivers entangled photons to the LTM ports of the QCs. In contrast, the second approach utilizes exciting pulses to generate entangled photons between neighboring QCs. Comparing the EPR generation and distribution at the source versus at both endpoints, the latter requires a greater number of point-to-point connections between the Exciting Pulse Generator and the QCs.

It is important to note that, unlike the case of EPR generation and distribution at the midpoint---where teleportation can occur between any pair of QCs---in the cases of EPR generation and distribution at the source or at both endpoints, teleportation is limited to neighboring QCs due to the mesh topology considered in this proposal. Therefore, to teleport a qubit between two non-adjacent QCs, a sequence of teleportations must be performed along a path of directly connected QCs.

\subsection{Impact on the Timing}
\begin{figure}
\centering
    \begin{minipage}{0.8\textwidth}
    \centering
    \includegraphics[width=1.0\textwidth]{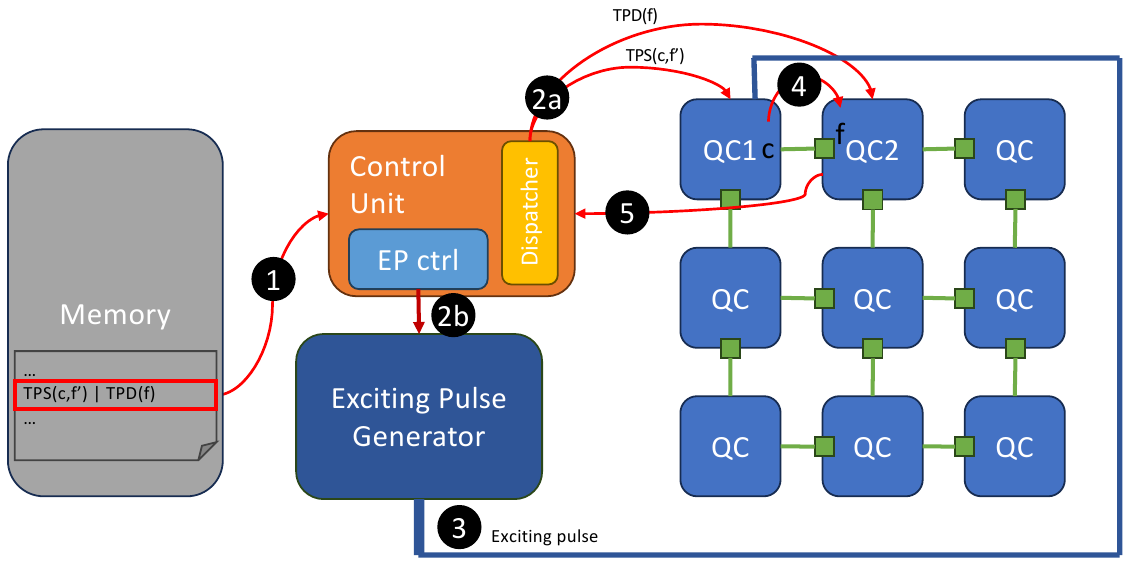}
    \subcaption{Phases involved in the execution of the remote bundle $\langle$\texttt{TPS(c,f) | TPD(f)}$\rangle$ when the EPR generation and distribution at source is used.}\label{fig:exemodel_source}
    \end{minipage}%
    \\
    \begin{minipage}{0.8\textwidth}
    \centering
    \includegraphics[width=1.0\textwidth]{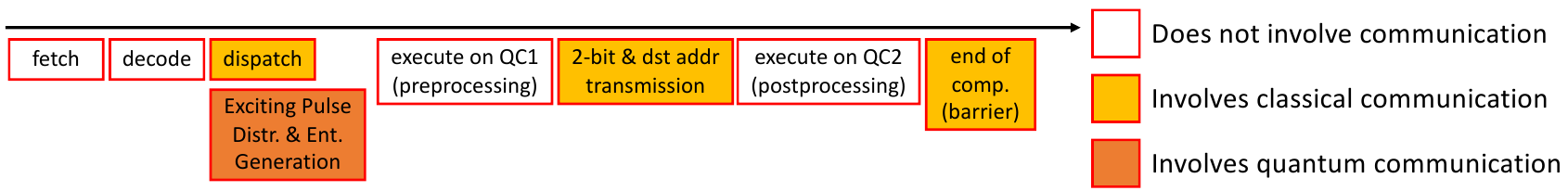}
    \subcaption{Execution timeline for a teleportation instruction when the EPR generation and distribution at source is used.}\label{fig:timing_source}
    \end{minipage}
\caption{Execution model and timeline for a teleportation instruction when the EPR generation and distribution at source is used.}
\end{figure}

The choice of a specific EPR generation and distribution technique does not impact either the execution model or the timing. This is illustrated in Figure~\ref{fig:exemodel_source}, which depicts the execution model and timing for a teleportation instruction between two neighboring QCs when EPR generation and distribution occur at the source. As shown, the execution steps align with those discussed for the case of EPR generation and distribution at the mid-point in Figure~\ref{fig:tpstpd}. The only difference is in step~3, where the Exciting Pulse Generator sends a pulse to the source node (QC1), which is responsible for generating the entangled EPR pair between QC1 and QC2. Notably, when using EPR generation and distribution at both endpoints, the Exciting Pulse Generators send pulses to both QC1 and QC2. 

The timing of the different steps is shown in Figure\ref{fig:timing_source}. As observed, it does not introduce significant variations compared to the case of EPR generation and distribution at the midpoint, discussed in Figure\ref{fig:timeline}. The only difference is the additional contribution from the exciting pulse distribution and entanglement generation, which occurs in parallel with the dispatch of instructions. Therefore, the timing model presented in Sec.~\ref{sec:timing_model} remains valid for all three EPR generation and distribution approaches.

\section{Other Examples}
\label{apdx:examples}
\begin{figure}
    \centering
    \includegraphics[width=0.99\columnwidth]{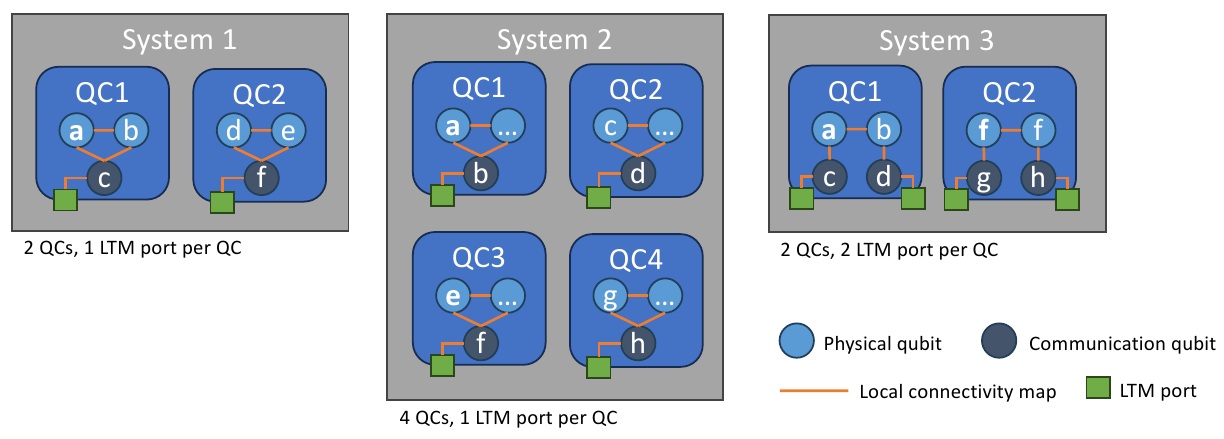}
    \caption{Circuit and considered system organizations.}
    \label{fig:sysorganization}
\end{figure}

In this section, we provide additional examples to illustrate how the assembly changes for the same circuit when the system organization varies in terms of the number of QCs and LTM ports per QC. We consider three different system configurations, as shown in Figure~\ref{fig:sysorganization}.

The circuit, involving four qubits, consists of a single slice with two CNOT gates. The first system configuration, System~1, includes two QCs, each equipped with one LTM port. The second configuration, System~2, consists of four QCs, each with one LTM port. Finally, the third configuration, System~3, features two QCs, each equipped with two LTM ports.

\subsection{System~1}
\begin{figure}
    \centering
    \includegraphics[width=0.9\columnwidth]{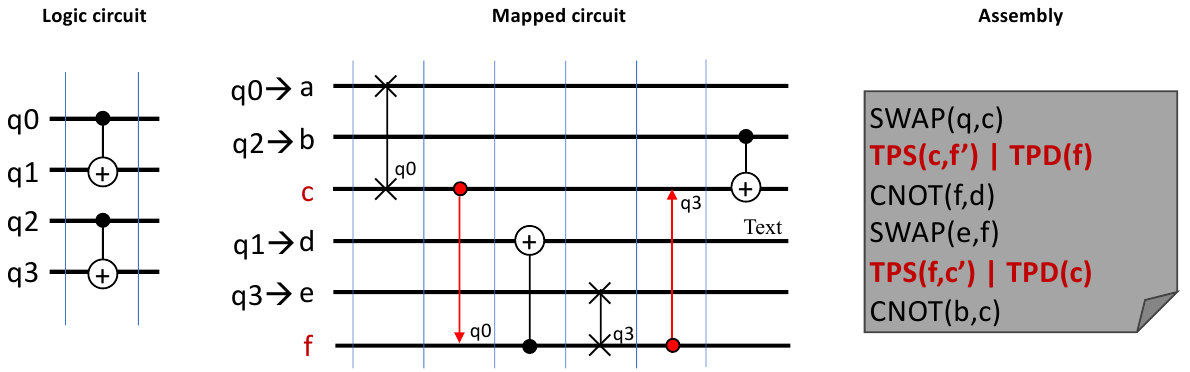}
    \caption{Logic circuit (left), mapped circuit based on organization System~1 (middle), and corresponding assembly code. The two teleportations are performed in sequence.}
    \label{fig:sys1}
\end{figure}
Figure~\ref{fig:sys1} illustrates the mapping of the 4-qubit circuit onto System~1, where logical qubits $q0$, $q1$, $q2$, and $q3$ are mapped to physical qubits $a$, $b$, $d$, and $e$, respectively. In this configuration, the two CNOT gates cannot be executed within the same slice because their qubits reside in different QCs. As a result, two teleportations are performed: transferring $q0$ to QC2 and $q3$ to QC1. The corresponding assembly code is shown on the right side of the figure. As observed, no parallelism can be exploited in this setup and the two teleportations are performed in sequence.

\subsection{System~2}
\begin{figure}
    \centering
    \includegraphics[width=0.9\columnwidth]{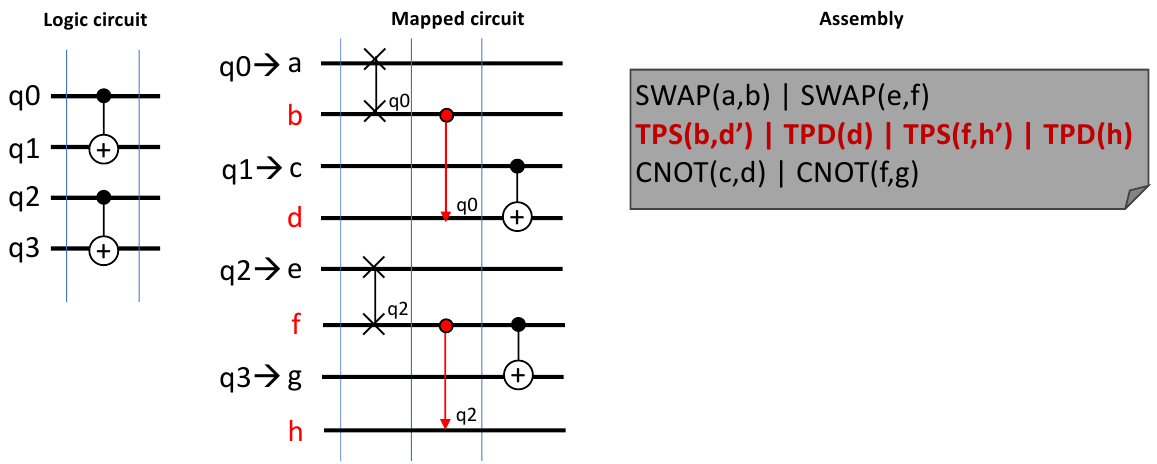}
    \caption{Logic circuit (left), mapped circuit based on organization System~2 (middle), and corresponding assembly code. The two teleportations are performed concurrently.}
    \label{fig:sys2}
\end{figure}

Figure~\ref{fig:sys2} demonstrates the mapping of the previously considered circuit onto System~2, where logical qubits $q0$, $q1$, $q2$, and $q3$ are assigned to physical qubits $a$, $c$, $e$, and $g$, respectively. Since the physical qubits involved in the two CNOT gates are located in different QCs, two teleportations are necessary to bring the relevant qubits into the same QC for each CNOT operation. In this example, $q0$ is transferred from QC1 to QC2, and $q2$ from QC3 to QC4. Notably, the two teleportations can be performed concurrently.

\begin{figure}
    \centering
    \includegraphics[width=0.9\columnwidth]{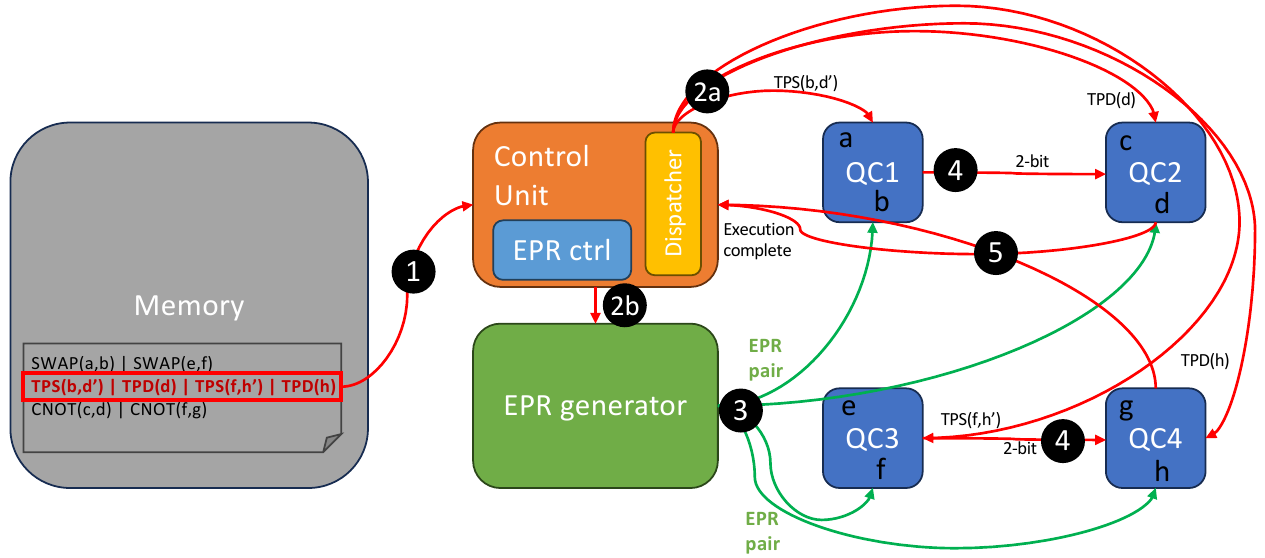}
    \caption{Phases involved in the execution of bundle $\langle$\texttt{TPS(b,d') | TPD(d) | TPS(f,h') | TPD(h)}$\rangle$.}
    \label{fig:sys2-execution}
\end{figure}

The sequence of phases performed for the execution of bundle $\langle$\texttt{TPS(b,d') | TPD(d) | TPS(f,h') | TPD(h)}$\rangle$ is shown in Figure~\ref{fig:sys2-execution}. Phase (1) is the fetch of the instruction from the memory. In phase (2) the \emph{dispatcher} dispatches the four instructions into the appropriate QCs and the \emph{EPR ctrl} configures the \emph{EPR generator} to generate two entangled EPR pairs that are delivered to the four QCs. In phase (3), the first EPR pair in transferred to QC1 and QC2 and the second to QC3 and QC4 and the pre-processing step of the teleportation protocol is performed in the source QCs, namely, QC1 and QC3. In pahse (4) the 2-bit of classical information are transmitted from QC1 to QC2 and from QC3 to QC4 and the post-processing phase of the teleportation protocolo is performed. Finally, in phase (5), QC2 and QC3 send an execution completion message to the \emph{control unit}.

The sequence of phases involved in executing the instruction bundle $\langle$\texttt{TPS(b,d') | TPD(d) | TPS(f,h') | TPD(h)}$\rangle$ is illustrated in Figure~\ref{fig:sys2-execution}. In phase (1), the instruction bundle is fetched from memory. During phase (2), the \emph{dispatcher} distributes the four instructions to the corresponding QCs, while the \emph{EPR ctrl} configures the EPR generator to create two entangled EPR pairs, which are delivered to the four QCs. Phase (3) involves transferring the first EPR pair to QC1 and QC2, and the second to QC3 and QC4, followed by the pre-processing step of the teleportation protocol in the source QCs, QC1 and QC3. In phase (4), the two-bit classical information is transmitted from QC1 to QC2 and from QC3 to QC4, enabling the post-processing phase of the teleportation protocol. Finally, in phase (5), QC2 and QC4 send an execution completion message to the \emph{control unit}.

\subsection{System~3}
\begin{figure}
    \centering
    \includegraphics[width=0.9\columnwidth]{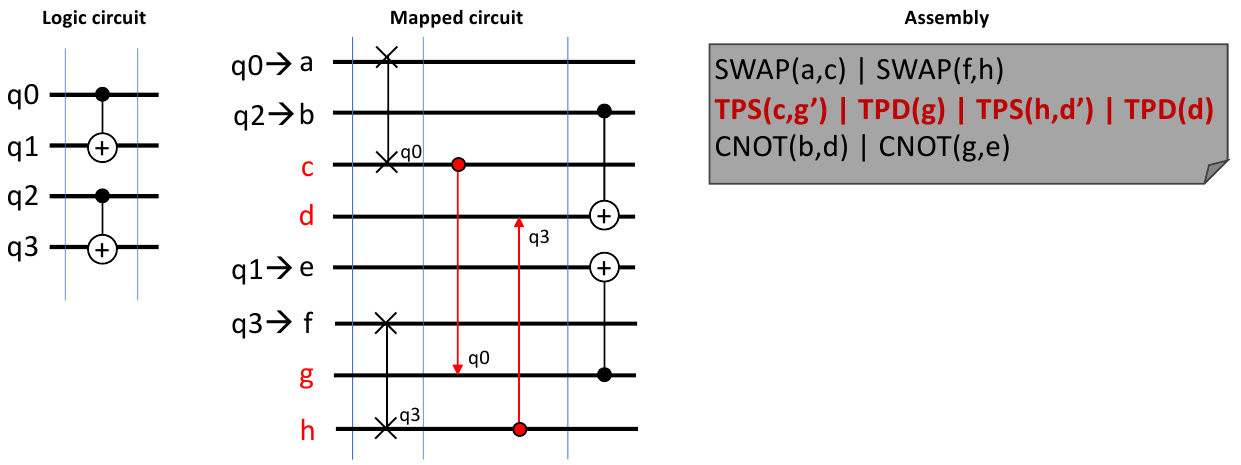}
    \caption{Logic circuit (left), mapped circuit based on organization System~3 (middle), and corresponding assembly code. The two teleportations are performed concurrently.}
    \label{fig:sys3}
\end{figure}

Figure~\ref{fig:sys3} illustrates the mapping of the same circuit onto System~3. Similar to System~2, two teleportations can be performed simultaneously. In this configuration, logical qubits $q0$, $q1$, $q2$, and $q3$ are mapped to physical qubits $a$, $e$, $b$, and $f$, respectively. To execute the two CNOT gates, $q0$ is transferred to QC2, and $q3$ is transferred to QC1. Since each QC has two LTM ports, these transfers can be carried out through two concurrent teleportations.

\begin{figure}
    \centering
    \includegraphics[width=0.9\columnwidth]{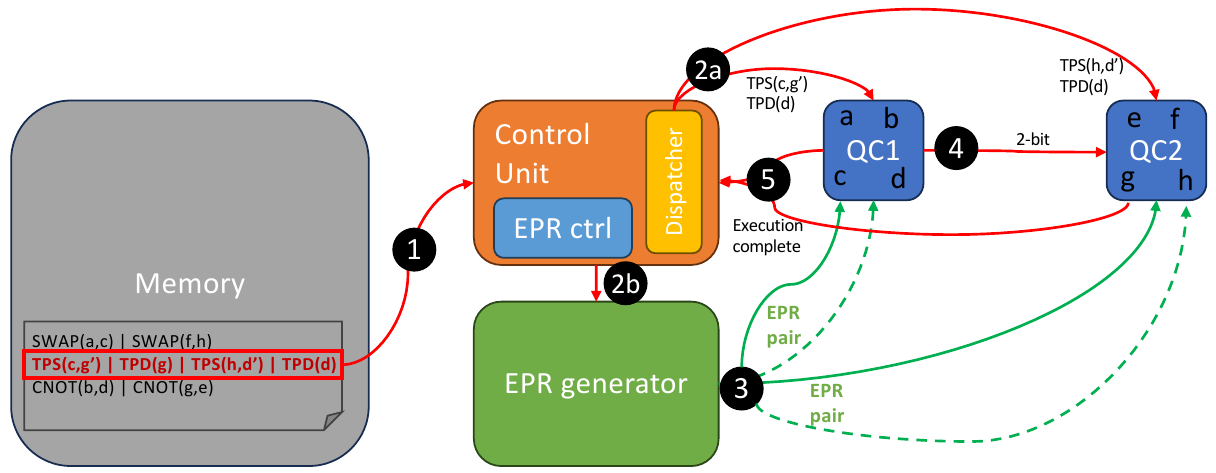}
    \caption{Phases involved in the execution of bundle $\langle$\texttt{TPS(c,g') | TPD(g) | TPS(h,d') | TPD(d)}$\rangle$.}
    \label{fig:sys3-execution}
\end{figure}

The phases involved in executing the bundle $\langle$\texttt{TPS(c,g') | TPD(g) | TPS(h,d') | TPD(d)}$\rangle$ are illustrated in Figure~\ref{fig:sys3-execution}. In phase (2), the \emph{dispatcher} sends instructions \texttt{TPS(c,g')} and \texttt{TPD(g)} to QC1, and \texttt{TPS(h,d')} and \texttt{TPD(d)} to QC2. During the same phase, the \emph{EPR ctrl} configures the \emph{EPR generator} to create two EPR pairs. In phase (3), the first EPR pair is delivered to the first LTM ports of QC1 and QC2, while the second EPR pair is sent to their second LTM ports (represented by solid and dashed lines, respectively). The remaining phases proceed similarly to the previous cases.

\section{Finite State Machine of the Control Unit}
\label{apdx:cu_fsm}
\begin{figure}
    \centering
    \includegraphics[width=0.95\columnwidth]{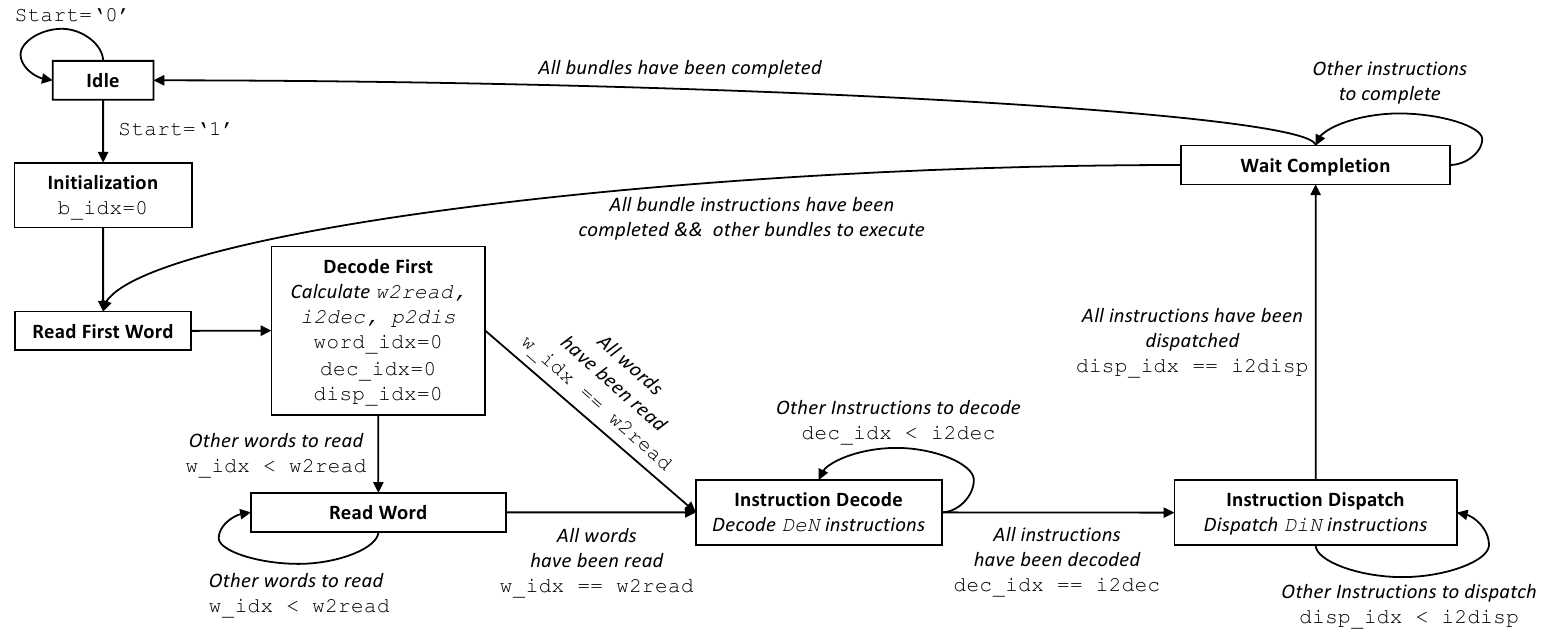}
    \caption{Finite state machine of the control unit.}
    \label{fig:cu_fsm}
\end{figure}

Figure~\ref{fig:cu_fsm} shows the finite state machine (FSM) of the CU.
When the program starts, the CU transitions from the \emph{Idle} state (inactive) to the \emph{Initialization} state, where the bundle execution counter and the number of words read per bundle are reset. Next, in the \emph{Read First Word state}, the system reads the first word, which contains the total number of instructions in the bundle along with the first $k$ instructions. The value of $k$ depends on the word size and the number of instructions that can fit within a single word. After reading the first word, the CU moves to the \emph{Decode First} state, where it calculates: i) the total number of words to be read (\texttt{w2read}) to acquire the complete bundle, ii) the number of instructions to be decoded (\texttt{i2dec}), and iii) the number of packets to be dispatched to the QCs (\texttt{p2dis}). If no additional words are required, the system proceeds directly to the \emph{Instruction Decode} state, where it decodes the instructions within the single word of the bundle. Otherwise, it transitions to the \emph{Read Word} state, reading the remaining words of the bundle before moving to the \emph{Instruction Decode} state to process all instructions. Since a bundle contains $N(B)$ instructions and up to \texttt{DeN} instructions can be decoded simultaneously, the decoding phase (\emph{Instruction Decode}) is performed iteratively, processing \texttt{DeN} instructions at a time. Once all instructions in the bundle have been decoded, the CU transitions to the \emph{Instruction Dispatch} state, where up to \texttt{DiN} instructions are dispatched, all directed to the same QC. After dispatching all instructions in the bundle, the CU enters the \emph{Wait Completion} state, awaiting confirmation of execution completion from the QCs. Once execution is complete---either because all instructions have finished processing or the CU has received completion messages from all QCs---the system moves to the next bundle, restarting from the \emph{Read First Word} state. If all bundles have been executed, the program terminates.

\end{document}